\documentclass[11 pt]{article}
\usepackage{t1enc}
\usepackage{amsfonts}
\input xy
\xyoption{all}
\def\be{\begin{eqnarray}}
\renewcommand{\le}[1]{\label{#1}\end{eqnarray}}
\def\ee{\end{eqnarray}}

\usepackage{amsmath, amssymb}
\usepackage[dvips]{graphicx}

\newtheorem{theorem}{Theorem}

\newcommand{\edth} {\mbox{\symbol{'360}}}
\begin{document}

\rightline{SPIN-03/18}
\rightline{ITP-UU-03/27}
\rightline{hep-th/0306??}
\rm
\vskip 0.3in

\begin{center}
{\Large \textbf{\ Exploring the holographic principle in asymptotically flat spacetimes via the 
BMS group}}

\vspace{1 cm}

{\bf {Giovanni Arcioni}}$^{a,}$\emph{\footnote{Address after October 2003: The 
Racah Institute of Physics, The Hebrew University, Jerusalem 91904, Israel.
}$^,$\footnote{
E-mail : G.Arcioni@phys.uu.nl} }and {\bf {Claudio Dappiaggi}$^{b,}$\footnote{
E-mail : claudio.dappiaggi@pv.infn.it},}

\vspace{1cm}

$^{a}$ {\it Spinoza Institute and Institute for Theoretical Physics,\\ Leuvenlaan 4,
3584 CE Utrecht, The Netherlands}

\vspace{0.1cm}

$^{b}${\it ~Dipartimento di Fisica Nucleare e Teorica,

Universit\`{a} degli Studi di Pavia, INFN, Sezione di Pavia, \\[0pt]
via A. Bassi 6, I-27100 Pavia, Italy}

\end{center}
\begin{abstract}

We explore the holographic principle in the context of asymptotically flat 
spacetimes. In analogy with the AdS/CFT scenario we analyse the 
asympotically symmetry group of this class of spacetimes, the so called 
Bondi-Metzner-Sachs (BMS) group. We apply the covariant entropy bound to 
relate bulk entropy to boundary symmetries and find a quite different 
picture with respect to the asymptotically AdS case. We then derive the 
covariant wave equations for fields carrying BMS representations to investigate 
the nature of the boundary degrees of freedom. We find some similarities with 't Hooft 
S-matrix proposal and suggest a possible mechanism to encode 
bulk data.

\end{abstract}

\newpage

\tableofcontents

\section{Introduction}

Some years ago 't Hooft \cite{salamfest} proposed that the apparent 
paradoxes of black hole physics in local quantum field theory can be 
resolved if the fundamental theory of quantum gravity has degrees of 
freedom living on lower dimensional hypersurfaces with respect to the 
bulk spacetime on which information is encoded ``holographically''.

This was motivated by the Bekenstein-Hawking formula for the black hole 
entropy $S_{BH}=A/4$, which says that the entropy
 goes like an area instead of a volume. It is thus commonly assumed that a holographic theory 
has a density of states equal to the Planck density. This implies an extreme 
reduction  in the complexity of a physical system and it is implicitly stated 
that gravity is responsible for this huge reduction of the number of degrees 
of freedom. In other words, the usual way of counting states in QFT is highly 
redundant since if we try to excite more than $A/4$ degrees of freedom we end up 
with the formation of a black hole.

The holographic principle predicts then a quite considerable
 departure from conventional way 
of counting degrees of freedom in physical systems. We can imagine two 
ways of producing a holographic behaviour. One possibility is to preserve 
locality at the price of a sort of (very!) unusual ``gauge symmetry'' which should 
be able to give the correct counting required by the holographic bound. This is essentially the recent 
approach advocated by 't Hooft \cite{dissipative}: one has
 ``ontological'' states which go as a volume as usual, while 
``equivalence classes'' grow as a surface and the latter are
 supposed to produce (this mechanism is still to be explained) 
 the area law in the case of black holes. This approach, of course, requires a 
reformulation of quantum mechanics itself.

Another option is to reconstruct spacetime starting from 
holographic data. In this case one has to explain how they are generated, their dynamics 
and how they can produce classical spacetime geometry. Locality is simply
 recovered a posteriori. AdS/CFT correspondence \cite{ADS/CFT} gives
  a beautiful example of
 this kind of behaviour. Nevertheless, 
how to 
recover bulk locality seems still an open question and most important the whole 
approach assumes an equivalence between partition sums of gravity and gauge theory 
once asymptotically AdS boundary conditions are imposed. As we will 
see in the following this is a very special choice and the fact that a  
conventional QFT appears on the boundary is in some sense quite unique. 

It is therefore interesting to see what happens in the case of different 
boundary conditions. In the present paper we simply explore from a 
general point of view the case of asymptotically 
flat spacetimes. As we will see there is a sharp difference with respect to 
the case of AdS boundary conditions and a unique (geometrical) reconstruction of spacetime 
seems unlikely, suggesting a somehow different way of encoding and storing 
bulk data. With the help of the covariant 
entropy bound \cite{holoreview} we analyse some aspects 
of the asymptotic symmetry group (ASG) 
of asymptotically flat spacetimes-the so called Bondi-Metzner-Sachs (BMS)
group- and the nature of the fields carrying its representations 
hoping to get some insights about a possible holographic 
description.
\begin{center}
{\it Organization of the paper}
\end{center}
In Section 2 we review the notion of asymptotic symmetry group and discuss 
its role and importance in the context of the holographic principle.

In Section 3 we recall different derivations of the BMS group and examine 
some of its properties and subgroups. We concentrate in particular on the Penrose derivation, 
since more convenient for our holographic purposes. This resum\'ee of the BMS 
group is also intended 
to give the reader some background and terminology for further applications.

In Section 4 we review and apply the covariant entropy bound proposed by Bousso \cite{holoreview} 
pointing out a connection between entropy production in the bulk and the 
corresponding symmetries of a candidate holographic description. We then 
compare the situation with asymptotically AdS spacetimes and 
underline differences.

In Section 5 we review the representation theory of the BMS group and 
show how to label different states. Details of BMS representation
theory are contained in Appendix (A.1), (A.2); most of this material is 
used to construct covariant wave equations. 

In Section 6 we construct indeed covariant wave equations for the BMS group using 
a general framework based on fiber bundle techniques. Comments on the possible
 meaning 
of the various little groups and their labels are also given at the end. 

In Section 7 we try to interpret the results of Section 6 and
 explore possible features and aspects of a tentative holographic 
 description. We note 
 some similarities with 't Hooft 
 horizon holography and suggest some further aspects to take into account for a 
holographic description.

We end up in Section 8 with some concluding remarks.


\section{Origin of the
 asymptotic symmetry group and its role in the holographic context} 

In absence of gravity the isometry group of (Minkowski) spacetime is the 
well known Poincar\'e group, namely the semidirect product of the Lorentz group and
translations. Once gravity is switched on, even if weak, the situation changes 
quite drastically: Lorentz transformations, i.e. the homogeneous part of the
Poincar\'e group, still make sense at each point, but the Poincar\'e group as a 
whole is not of course any more an isometry of the manifold and  ``seems to
vanish into thin air'' \cite{sachsa}.

One might think to use as a suitable generalization the general coordinate 
transformation group, in other words diffeomorphisms. However, they simply 
preserve the differentiable structure (the ``smoothness'') of the manifold and they
are not so relevant from the physical point of view. 

One therefore relies on the asymptotic symmetry group (ASG). Roughly 
speaking, this means that one adds a conformal boundary to the original
 manifold following Penrose prescription and
considers asymptotic symmetries as conformal motions on the boundary preserving
some structure defined on it. Of course, these symmetries need not to be
extended to the bulk in general but they can play a role in the 
holographic reconstruction of spacetime. An explicit example is given in the 
AdS/CFT correspondence \cite{ADS/CFT}, where one ``reconstructs'' the bulk in the case 
of asymptotically AdS spacetimes \cite{skenderis}. This represents however a special case as 
we will see in the following, despite the common non-compactness of the boundary as 
in the case of asymptotically flat spacetimes.

It is important to stress that one is {\it free} to choose the 
conformal boundary, the choice being just a matter of 
convenience. In the set up proposed by Bousso \cite{holoreview}, this means that there 
are many different ways to project the bulk into collections of ``holoscreens''. Note 
also the it seems
natural to put the screen on the conformal boundary at infinity, but in principle 
{\it any} other choice is allowed. Again, from this point of view, AdS is a very special 
case, since there is a natural screen on the boundary of spacetime. We are going 
to consider asymptotically flat spacetimes
 and imagine the screen to 
be the null manifold $\Im$ at infinity but as said other choices can 
also be done in principle. One could also imagine to put the screen at spatial infinity, where there is 
some similarity to the null case from the point of view of the ASG (See comments at the 
end of Section 6).

The notion of ASG seems then particularly natural in the context of gravity once one goes 
beyond classical General Relativity: non local aspects, ``hidden'' at the classical 
level by choosing initial conditions for the solutions, should manifest 
themselves in the quantum regime, demanding and depending on 
the choice of boundary conditions.
\section{Emergence of the BMS group in asymptotically flat spacetimes}
In this section we briefly revisit the derivation of the BMS group and some of its 
properties. As said, the BMS group represents the ASG of asymptotically 
flat spacetimes. More properly it is a transformation pseudo-group of 
asymptotic isometries of the region close to infinity of the asymptotically 
flat (lorentzian) spacetimes \footnote{For the euclidean case see section 6.}. There is however a derivation proposed by Penrose as a group of transformations 
living intrinsically {\it on} $\Im$. In this case the BMS group {\it is} 
the transformation group on the boundary $\Im$.\footnote{Recall that $\Im$ is the 
disjoint union of $\Im^+$ (future null infinity) with $\Im^-$ (past null
infinity). In the rest of the paper we will refer to $\Im^+$ but because of the
symmetry 
the same conclusions will hold on $\Im^-$  too in all cases 
unless differently specified.}

>From the point of view of the holographic principle we are interested in the 
theory living on the boundary and its symmetry group. We therefore 
prefer to consider this ``boundary description'' of the BMS group as more relevant 
and fundamental for our purposes. Moreover, it keeps into 
account the degenerate nature of $\Im$, which one has to face up when 
choosing a null screen. For completeness, however, we review various derivations.
\subsection{BMS as asymptotic symmetry group}
The BMS group was originally discovered \cite{Bondi}, \cite{sachsa} by studying gravitational radiation 
emitted by bounded systems in asymptotically flat spacetimes; it is the 
group leaving invariant the asymptotic form of the metric describing 
these processes. Quite generally, one can choose $(u,r,\theta,\phi)$ coordinates 
close to null infinity and check that the components of the metric tensor behave 
like those of the Minkowski metric in null polar coordinates in the limit 
$r \rightarrow \infty$.

A BMS transformation $(\alpha,\Lambda)$ is then
\begin{gather} 
\bar{u} =[K_\Lambda (x)]^{-1} (u+ \alpha(x)) + O(1/r) \\
\bar{r} = K_\Lambda(x) r +J(x,u) +O(1/r)\\ 
\bar{\theta}= (\Lambda x)_\theta + H_\theta (x,u) r^{-1} + O(1/r)\\
\bar{\phi}=(\Lambda x)_\phi + H_\phi (x,u) r^{-1} +O(1/r)
\end{gather}

\noindent where $x$ is a point on the two sphere $S^2$ with coordinates $(\theta,\phi)$, 
$\Lambda$ represents a Lorentz transformation acting on $S^2$ as a conformal 
transformation and $K_\Lambda(x)$ is the corresponding conformal factor.
Furthermore $\alpha$ 
is a scalar function on $S^2$ associated with the so called supertranslation 
subgroup. It represents the ``size'' of the group as we will see below. 

The other functions are uniquely determined by $(\alpha,\Lambda)$ imposing
\begin{gather}
(\alpha_1,\Lambda_1)(\alpha_2,\Lambda_2)=(\alpha_1+\Lambda_1\alpha_2, \Lambda_1
\Lambda_2)\label{semid}\\
(\Lambda_1\alpha_2)(x)=[K_\Lambda (x)]^{-1} \alpha_2(\Lambda_1^{-1}x)\label{groupstr}
\end{gather}

\noindent One immediately notices from (\ref{semid}) the structure of semidirect product. Therefore the 
BMS group $B= N \ltimes L $ is the semidirect product of the infinite 
dimensional supertranslation group $N$ with (the connected component of the
homogeneous) Lorentz transformations group $L$.\footnote{We will consider
$SL(2,\mathbb{C})$, the universal covering group of $L$.}

An important point to keep in mind is that the ASG thus defined is {\it universal} 
since one gets the same group for all asymptotically flat spacetimes. This is 
quite surprising. In addition, the group is infinite dimensional due to the presence 
of extra symmetries which reflect the presence of gravity in the bulk.

It is also possible \cite{tamburino} to derive the BMS group B working in the 
unphysical spacetime and imposing differential and topological requirements 
on $\Im$, avoiding then asymptotic series expansions. In 
a sense, this is a finite version of the original BMS derivation, since 
one constructs a so called conformal Bondi frame in some finite
neighborhood of $\Im$ and this finite region corresponds to an infinite region 
of the original physical spacetime.

Even if one is not working with an asymptotic expansion, we prefer to 
consider this derivation from a slightly different perspective with respect to 
the one of Penrose, who considers the emergence of the BMS working 
intrinsically on $\Im$. Actually, 
we are still working asymptotically, 
 even if, in the unphysical space, 
 the ``infinite is brought to finite''. Recall, however, that had we chosen another 
conformal frame we would have obtained an isomorphic group. In other words, covariance 
is preserved by differentiable transformations acting on $\Im$. This indeed 
motivated Penrose to work directly on the geometrical properties of
$\Im$ as we will see below and it seems more convenient to investigate the 
holographic principle in this context.

Choosing $x^0=u$, $x^A=(\theta,\phi)$ as coordinates on $\Im^+$ and $x^1=r$ 
defining the inverse luminosity distance, the (unphysical) metric $g^{\mu \nu}$ in the conformal Bondi frame
is thus
$$g^{\mu \nu}=\left[\begin{array}{ccc}
0 & g^{01} & 0\\
g^{01} & g^{11} & g^{1A}\\
0 & g^{1A} & g^{AB}
\end{array}\right]$$ 
\noindent Using the freedom of gauge choice of the conformal factor 
 and imposing global and asymptotic requirements on $\Im$ one can write
the metric (in a neighbourhood of $\Im$) as 
$$g^{\mu \nu}=\left[\begin{array}{ccc}
0 & 1 & 0\\
1 & 0 & 0\\
0 & 0 & q^{AB}
\end{array}\right]$$ 
\noindent where $q^{AB}$ is the metric on the $S^2$ therefore time independent.
One can eventually compute the generators $\xi^\mu$ of asymptotic infinitesimal transformations 
$x^\mu \rightarrow x^\mu +\xi^\mu$
by solving 
\be
\xi^{(\mu ; \nu)}- (\Omega,_\rho \xi^\rho / \Omega) g^{\mu \nu} =0.
\ee 
One finds 
\begin{gather}
\xi^A=f^A(x^B)\\
\xi^0 = \frac{1}{2}u f_{;A}^A+ \alpha (x^B),\; \xi^1 =0.
\end{gather}
\noindent Setting the supertranslations $\alpha(x^A)$ to zero we get the
(orthochronous) 
Lorentz group while setting the $f^A$ to zero we get the group of supetranslations as 
expected.

These are exactly the Killing vectors found by Sachs \cite{sachsa} in studying radiation at null
infinity. However, one interprets
 the notion of asymptotic symmetry as follows: one declares an
infinitesimal asymptotic symmetry to be described by a vector field $\xi^a$ 
(more precisely an equivalence of vector fields in the physical spacetime) 
such that the Killing equation $L_\xi g_{\mu \nu} =0$ is satisfied to as 
good an approximation as possible as one moves towards $\Im$.

One can also consider another derivation of the BMS group proposed by 
Geroch \cite{geroch}. In a nutshell, this procedure considers the ASG as the 
group of consometries of $\Im$, i.e. conformally invariant 
structures associated with $\Im$. More properly, these
structures live on the so called 
``asymptotic geometry'', a 3 dimensional manifold diffeomorphic to $\Im$ endowed with a 
tensor structure. We prefer to consider as truly intrinsic the derivation 
of Penrose which we discuss below. Note however that Geroch approach has received 
a lot of attention and has been adopted in particular by Ashtekar and (many) others to endow 
$\Im$ with a symplectic structure to study then fluxes and angular momenta of 
radiation at null infinity.
\subsection{Penrose derivation of the BMS group}
The derivation of Penrose is based on conformal techniques and the null nature 
of $\Im$ is explicitly taken into account. The underlying idea is the 
following: consider a motion in the physical spacetime; this will naturally 
generate a motion in the unphysical spacetime and in turn a conformal motion 
on the boundary. The latter can persist, even if the starting physical spacetime 
has no symmetry at all providing thus a definition of ASG. However the degenerate 
metric on $\Im$ does not itself endow sufficient structure to define
 the BMS group.\footnote{More precisely, one considers a future/past 
3 asymptotically simple spacetime, with null $\Im^+$,$\Im^-$ and strong
asymptotic Einstein condition holding on it \cite{nonlinear}.} 

The natural structure living on $\Im$ is that of a inner (degenerate) conformal 
metric, the topology being $\mathbb{R} \times S^2$; the ``$\mathbb{R}$'' represent the 
null geodesic $\Im$ generators with ``cuts'' given by two dimensional spacelike
hypersurfaces each with $S^2$ topology. Choosing a Bondi coordinate
 system \cite{nonlinear}, one can indeed write the degenerate 
 metric on $\Im^+$ as
\be \label{degmetric}
ds^2=0.du^2+d\theta^2+\sin^2\theta d\phi^2.
\ee
\noindent Using stereographic coordinates for the two sphere
 $(\zeta=e^{i\phi}cot(\theta/2))$  one has
\be
ds^2=0.du^2+4d\zeta d \bar{\zeta}(1+\zeta \bar{\zeta})^{-2}.
\ee 
\noindent and recalling that all holomorphic
 bijections of the Riemann sphere are of the 
form
\be
f(\zeta)=\frac{a\zeta +b}{c \zeta +d}
\ee
\noindent with $ad-bc=1$, one immediately concludes that the metric (\ref{degmetric}) is preserved
under the transformations
\begin{gather} \label{nugroup}
u \rightarrow F(u,\zeta , \bar{\zeta}) \\
\zeta \rightarrow \frac{a\zeta +b}{c \zeta +d}
\end{gather}
\noindent These {\it coordinate} transformations define the so called
Newman-Unti (NU)
group \cite{nugroup}, namely the group of non reflective motions of $\Im^+$ preserving its 
intrinsic degenerate conformal metric. 

The NU is still a very large group. One can restrict it by requiring to preserve 
additional structure on  $\Im$. One actually enlarges the notion
of angles and endows $\Im$ with ``strong conformal geometry''. In addition to ordinary 
angles one considers null angles: finite angles are formed by two 
different directions in $\Im$ at a point in $\Im$ which are not coplanar with the 
null direction in $\Im$, while null angles are formed by two directions at a 
point in $\Im$ which are coplanar with the null direction.

One can show that the set of strong conformal geometry preserving transformations 
restricts the $u$ transformations of the NU group to the following form
\be
u \rightarrow  K(u+ \alpha(\zeta,\bar{\zeta}))
\ee
\noindent with
\be
K=\frac{1+\zeta \bar{\zeta}}{(a\zeta+b)(a^* \zeta^*+b^*)+(c 
\zeta +d)(c^* \zeta^* +d^*)}
\ee
\noindent the same appearing in (\ref{groupstr}) and $\alpha(\zeta,\bar{\zeta})$ an
arbitrary function defined on the two
sphere. But then this set of transformations together with the conformal transformations 
on the two sphere define precisely the BMS group as shown before. 

Note that in terms of this intrinsic description, these transformations have to be interpreted 
not as coordinate transformations but as {\it point} transformations mapping 
$\Im$ into itself. In other words, a conformal transformation of $\Im$ induces 
conformal transformations between members of families of asymptotic
2-spheres when moving 
along the affine parameter $u$. 

This construction further motivates the mapping between $\Im$ and the so called 
cone space\cite{bramson}, which we are going to discuss in the following as a possible abstract 
space where the holographic data might live.

Finally one has to remember that the global structure of
the BMS group in {\it four} dimensions cannot be generalized to a generic dimension as
$BMS_d=N_{d-2}\ltimes SL(2,\mathbb{C})$ where $N_{d-2}$ is the abelian group of scalar
functions from $S^{d-2}$ to the real axis; an example is the three dimensional
case
where \cite{Ashtekar3} it has been shown that $BMS_3=N_1\ltimes Diff(S^1)$. In what 
follows we are always going to work in {\it four} dimensions.

\subsection{BMS subgroups and angular momentum}

We review a bit more in detail the BMS subgroups. One 
has the subgroup N given by
\begin{gather} \label{supertr}
u \rightarrow u+ \alpha(\zeta,\bar{\zeta}) \\
\zeta \rightarrow \zeta
\end{gather}
\noindent known as supertranslations. It is an infinite
dimensional abelian subgroup; note that 
\begin{equation}
\frac{ BMS}{N}\simeq  SO(3,1)\simeq
PSL(2,\mathbb{C}),
\end{equation}
which follows from the fact that the BMS group is the semidirect product of N 
with SL(2,C). Choosing for the conformal 
factor $K$ on the sphere
\be
K=\frac{A+B\zeta +B^* \bar{\zeta} +C \zeta \bar{\zeta}}{1+\zeta \bar{\zeta}}
\ee
one has the subgroup $T^4$ of translations which one can prove to be 
the unique 4-parameter normal subgroup of N.

On the other hand, the property of a supertranslation to be translation 
free is {\it not} Lorentz invariant. Therefore there are several Poincar\'e groups at 
$\Im$, one for each supertranslation which is not a translation, and none of
them is preferred. This causes the well known difficulties in asymptotically 
flat spacetimes in defining the angular momentum, the origin basically being 
``free'' (because of the presence of gravity). We recall for completeness the reason: in Minkowski space-time, the angular momentum is described
 by a skew symmetric tensor
which is well-defined up to a choice of an origin. Whereas this last condition
is equivalent to fix 4 parameters, in the case of $\Im$ this condition requires 
an infinite number of parameters to fix the ``orbital'' part of the 
angular momentum  since the translation group $T^4$ is
substituted by the supertranslations N.

 Although many ways to circumvent this
problem have been proposed, no really satisfactory solution has emerged until
now. In \cite{Ashtekar2}, \cite{penrose} one ends up with a reasonable 
definition of
angular momentum in asymptotically stationary flat space-times, where
the space of good cross sections (i.e. sections with null asymptotic shear) is not empty; one can 
then select a Poincar\'e subgroup from the BMS group and define accordingly the angular momentum. We are going to examine 
in more detail the notion of good/bad sections in the following so as to discuss bulk 
entropy production from the point of view of boundary symmetries.

\section{Bulk entropy and boundary symmetries}

\subsection{Bousso covariant entropy bound}

We start with a very brief review of Bousso covariant entropy bound. We refer to \cite{holoreview}
for more details and examples. Here we just recall the 
relation between entropy and focusing of light rays examining then
this link from the point of view of BMS symmetries acting on null infinity.

Bousso bound represents a covariant generalization 
of the well known Bekenstein bound, the latter being 
a sort of spacelike version of the former. According to this more general
recipe, the entropy of matter that flows through lightsheets associated with a 
given two dimensional spacelike surface in spacetime is bounded by the area 
of that surface.

Bousso bound holds under some assumptions and it can be in principle violated by quantum effects. It therefore 
gives entropy estimates at the classical level. Nevertheless, it provides a general 
formulation of the holographic principle and one can construct ``screens'' 
on which the entire bulk information can be projected and stored.

Lightsheets play a fundamental role in this set up. This is not 
surprising, since General Relativity at the classical level
 can be thought as a series of 
``lenses'' filling spacetime! We recall indeed that once the two dimensional spacelike surface A 
has been chosen one defines lightsheets of A, L(A), as a null hypersurface that is 
bounded by A and constructed by following families of light rays orthogonally 
away from A, such that the cross sectional area is everywhere decreasing or 
constant. 
If S represents the entropy ``inside'' on any one of its L(A), then the bound 
predicts S less or equal to A/4.

One follows the light rays till they end up with a caustic in spacetime, i.e. the 
expansion $\theta$ goes from minus infinity to plus infinity. The 
change of expansion $\theta$ along a congruence of light rays is described by 
Raychaudhuri equation
\be
\frac{d\theta}{d\lambda}= - \frac{1}{D-2} \theta^2 - \sigma_{ab} \sigma^{ab}
+\omega_{ab}\omega^{ab}- 8 \pi T_{ab} k^a k^b
\ee
Since the twist vanishes for L and the last term will be non positive assuming
null energy condition if follows that the r.h.s is never positive and by solving
\be
\frac{d\theta}{d\lambda} < - \frac{\theta^2}{D-2}
\ee
one gets to the focusing theorem predicting an infinite value of the expansion
at the caustics due to bending light matter.

So energy costs
entropy and the latter focuses light and allows for the formation of 
caustics. More entropy crosses the lightsheets, faster the lightsheets end up. A system
in which light rays end up with some sort of percolation or random walk has therefore
more entropy than a system where they terminate in a point. A spherically symmetric 
system will have less entropy
then a system in which inhomogeneities are present: lightsheets in the second case 
will have to probe small scale density fluctuations and these spacetime irregularities 
will be associated with more entropy.

In the following we would like to interpret bulk irregularities which
are responsible for entropy production from the point of view of BMS boundary symmetries, 
identifying explicitly those which correspond to low/high entropic bulk 
configurations. 

Before doing this we would like to observe that the screen construction
that one can eventually activate following Bousso approach -i.e. the projection and
the storing of bulk information on suitable screens- is background dependent and 
also depends on the asymptotic structure of spacetime. As already mentioned before, 
choosing an asymptotic structure of spacetime is a matter of convenience, 
General 
Relativity being perfectly defined by itself.

\subsection{Bulk entropy and BMS boundary symmetries}
As previously recalled the BMS group is defined as those mappings acting on 
$\Im$ which preserve both the degenerate metric and the null angles.\\
In the case of null infinity, one can associate
 a {\it complex} function 
$\sigma(r,u,\zeta,\bar{\zeta})$, which in physical terms is a measure of the 
shear of the null cones which intersect $\Im^+$ at constant $u$. To define the 
shear one chooses a spinor field $O^A$ whose 
flagpole directions point along the the null geodesics of the congruence. The 
complex shear $\sigma$ is then defined as follows
\be
O^A O^B \nabla_{AA'} O_B = \sigma \bar{O_{A'}}
\ee
The argument of the shear $\sigma$ defines the plane of maximum shear and its modulus 
the magnitude of the shear. Now in the case of mild divergence
 of null geodesics (as with the Bondi-type hypersurfaces we are considering) one has
\be
\sigma = \frac{\sigma^0}{r^2} + O(\frac{1}{r^3})
\ee
The quantity $\sigma^0$ is $r$ independent and it is a measure of
 the {\it asymptotic} 
shear of the congruence of null geodesics intersecting $\Im^+$ at constant $u$. The 
$r$ independence is in agreement with the peeling-off \cite{rindler} properties of the 
radiation.

One can also read the shear from the asymptotic expansion of the metric. Consider 
for example the metric originally proposed by Sachs \cite{sachsa} 
\be
ds^2= e^{ 2 \beta} V r^{-1} du^2 -2 e^{2 \beta} dudr+r^2 h_{ab}(dx^a-U^a du)(
dx^b-U^bdu)
\ee
with $a,b$ indices running over 
angular coordinates and $V,\beta,U^a, h_{ab}$ are functions of 
the coordinates $(u,r,\theta,\phi)$ to be expanded in $1/r$ powers.
 The shear appears in the expansion of the function $\beta$
\be
\beta = -\sigma(u,\zeta,\bar{\zeta}) \sigma^* (2r)^{-2}+O(r^{-4})
\ee

Now when $\sigma^0 =0$, i.e. the asymptotic shear of the congruence of null 
geodesics vanish at infinity, one has ``good'' cross sections. On the other
hand, when non vanishing, one has ``bad'' sections. The latter corresponds to 
null geodesics ending up with complicated 
crossover 
regions in the bulk. Good cross sections do not exist in general spacetimes. However, a very 
special situation occurs in stationary spacetimes: in this case one can find
asymptotic shear free sections and the space of such cuts is isomorphic to Minkowski space
time, where a good section corresponds trivially to the lightcone originating 
from a point in the bulk. Of course  in the case of stationary spacetimes points 
of the isomorphic Minkowski space are not in one to one correspondence with 
points of the physical curved spacetime; the behaviour in the bulk of 
null geodesics will be however quite mild (compared to 
bad sections) to end up in an almost clean 
vertex. 

The intersection of the congruence of null geodesics originating from the bulk
 with $\Im$  is a connected two dimensional spacelike surface so we
  can apply the covariant entropy bound and  
  deduce that bad cross sections will in general correspond to more entropic 
configurations from bulk point of view. Indeed  in the case of 
bad cross sections lightrays ``percolate'' more than 
in the case of good sections, producing therefore more entropy.

One can say more and relate the notion of good/bad cuts to BMS boundary symmetries, having 
in mind a tentative holographic description. Indeed,
 under BMS supertranslations the transformation rule among asymptotic 
shears is
\be
\sigma^{o'}(u',\zeta,\bar{\zeta})=\sigma^o (u'-\alpha,\zeta,{\bar{\zeta}}) +(\edth)^2 
\alpha(\theta, \phi)
\ee
where the operator $\edth$ on the r.h.s is the so called ``edth'' operator (for a definition see
\cite{rindler}). One is then interested in finding transformations which produce new good
cuts. For Minkowski spacetime and (remarkably!) again stationary spacetimes one can map good cuts
into good cuts by means of translations and in these cases the BMS group can
be reduced to the Poincar\'e group by asking for the subgroup of the BMS transformations
which map good cuts into good cuts. In the general case, however, there are 
no good, i.e. asymptotic shear free sections, and from
BMS point of view this corresponds to not Lorentz 
free supertranslations. 

Applying therefore the covariant entropy bound one finds that bad sections correspond to more entropic 
configurations in the bulk and (not Lorentz free) supetranslations on the 
boundary. Time dependence produces more irregularities in the bulk 
giving therefore more entropy according to previous considerations; this is 
interestingly reflected in complicated supertranslations acting on the null
 boundary. 
 
If holographic data are stored then in the
  $S^2$ spheres on $\Im$ some of them will contain more/less information corresponding 
 to more/less entropy in the bulk. We return to this point in Section 7.

As said, asymptotic vanishing shear allows to reduce the BMS goup to Poincar\'e. One might 
think to start from the stationary case then for simplicity. There is still 
however a remnant of supertranslations. Suppose indeed to consider a system which 
emits a burst of radiation and it is stationary before and after the burst. The 
corresponding Poincar\'e subgroups will be different and they will have in common 
{\it only} their translation group. They will be related by means of a non trivial 
supertranslation in general. This is quite different from what 
happens in the AdS case as we are going to see in the next Section. 

\subsection{Difficulties in reconstructing spacetime and comparison with 
the AdS case} 
We now continue the previous analysis and make some comparisons with the 
AdS case.

Let us consider again the asymptotic shear. The previous picture tells 
us that the ASG can be reduced to Poincar\'e in some specific points along 
the boundary where the asymptotic shear does indeed vanish. This means that small 
shapes are preserved asymptotically as we follow the lines generating the null 
congruence using to construct the lightsheets. However, lightsheets acquire
in general shear in the asymptotic region. This is due tidal forces which
are responsible for the bending of light rays. But these are 
in turn described by the (asymptotic)
Weyl tensor and  this quantity (more properly the 
rescaled one in the unphysical metric) enters into the 
definitions of the so called Bondi news functions \cite{sachsa} which measure the amount 
of gravitational radiation at infinity. There is however another tensor, namely the Bach 
tensor (recall we are in four dimensions) $B_{\alpha \beta}$ which does not vanish in the presence of non zero Bondi news. In the case 
of asymptotically flat spacetimes it is {\it not} zero asymptotically. This is 
in {\it sharp} contrast with asymptotically AdS cases, where it vanishes
 asymptotically, the 
Bondi news being zero in that case. Actually the condition 
$B_{\alpha \beta} =0$ on 
$\Im$ {\it is} used in the definition \cite{magnon} of asymptotically AdS spacetimes.

This has however deeper consequences for holography. In AdS case this allows to 
reduce (enormously) the diffemorphism group on the boundary precisely to the conformal 
group. There is then (as already noticed in \cite{magnon}) a discontinuity in taking 
the limit $\Lambda\rightarrow0$ of the cosmological constant. In 
asymptotically flat spacetimes it means that 
one cannot propagate the boundary data to reconstruct the bulk in a {\it unique}
 way.
And this was the essence of the Fefferman-Graham theorems for the AdS case
\cite{fefferman}. 
We therefore see a remarkable difference. As a consequence, it
 also seems quite unlikely that 
GKPW (Gubser-Klebanov-Polyakov-Witten) prescription 
relating bulk-boundaries partition functions 
holds in this case. It seems also difficult to recover a S-matrix for 
asymptotically flat spacetimes starting from AdS/CFT and taking then the 
large radius limit of AdS.

As observed before, in general backgrounds the asymptotic shear does not vanish and therefore 
we remain with a big group on the boundary. Notice
 that using relativistic generalization 
of Navier-Stokes theory one can show that precisely the Bach current \cite{glass}
 can be used to describe 
entropy production in the bulk in the 
case of non stationary spacetimes. We see therefore that all the times (basically the majority) we cannot
 reduce BMS supertranslations to translations we have 
more entropy production in the bulk according to the covariant entropy bound and 
the production of this entropy can be measured in a quantitative way 
just by using the Bach current, a quantity which translates the effect of the 
bending of light before the system reaches equilibrium.

The fact that boundary symmetries cannot be reduced as in AdS case suggests not only that 
the propagation in the bulk is not unique (therefore we don't see the possibility of 
a naive holographic RG) as in AdS case but also that a degree of non 
locality\footnote{The emergence of non locality in asymptotically flat spacetimes
can also
be  explained considering again light propagation. Indeed, in AdS/CFT
the amount of time required for a light ray to cross
AdS diagonally is equal to the amount of time the light ray needs to "
go around the boundary". This is not true in  flat space. Therefore it
seems  unlikely that the holographic dual to asymptotically flat
spacetimes will be a local theory.} will be present-because of the impossibility of reducing 
the big symmetry group in general- in the candidate 
boundary theory, where fields will carry in general 
representations of the BMS group.

This motivates the following analysis of 
wave equations for the BMS group.

\section{Representations of the BMS group}    

As said our target is to write (covariant) wave equations as commonly done 
in physics for other groups \cite{Group2}. We therefore first review  very 
briefly in this Section the 
representation theory of the BMS group (See Appendix A.1 and A.2 
for details). We recall the situation for the Poincar\'e group to compare then similarities and 
differences with respect to the BMS case. We give the ``kets'' 
 to show explicitly the labelling of the corresponding states. Theory and 
 definitions 
 used in induced representations of semidirect product groups are reviewed 
 in Appendix A.

\begin{center}
{\large Poincar\'e group}
\end{center}

In this case we deal
with $P=T^4 \ltimes SL(2,C)$ whose little groups and orbits are well known and are summarized in the following table (see \cite{Simms} and
\cite{Group})\footnote{The Lie algebra of the 2-d Euclidean group
is:
$$[L_3,E_\alpha]=i\epsilon_{\alpha\beta}E_\beta,$$
$$[E_\alpha,E_\beta]=0.$$
The two Casimirs of the group are $E^2$ and $C_2=exp(2\pi iL_3)$ where $C_2=\pm
1$ (integer and half-integer values of $L_3$). The two $E(2)$ are actually the same group with different
representations depending if the value of the Casimir operator $E^2$ is
different (first case) or equal (second case) to zero}:

\vspace{0.5cm}

\begin{tabular}{|c|c|c|} 
\hline
Little group & orbit invariant & representation label \\
\hline
$SU(2)$ & $p^2=m^2$, sgn($p_0$) & discrete spin $j$ (dim=2j+1)\\
\hline
$SU(1,1)$ & $p^2=-m^2$, & discrete spin $j^\prime$\\
\hline
$E(2)$ & $p^2=0$, sgn($p_0$) & $\infty$-dimensional,\\
\hline
$E(2)$& $p^2=0$, sgn($p_0$) & 1-dimensional $\lambda$. \\ 
\hline
\end{tabular}

\vspace{0.5cm}

We first notice that the
generators of $T^4$ can be simultaneously diagonalized and for this reason the
orbit is the spectrum of energy. We have to impose some physical
restrictions, namely we call unphysical those representations related to negative
square mass and negative sign of $p_0$. Unfortunately, this is not enough since
we have to deal with a continuous
spin coming from $E(2)$. This case is excluded by hand and so we end up with
two spin quantum numbers, i.e. $j$ from $SU(2)$ and $\lambda$ from $E(2)$;
therefore the general ket for the Poincar\'e group is 
\begin{equation}\label{ketp}
\mid {\bf p}, j>,\;\;\;\mid {\bf p},\lambda>,
\end{equation}
respectively for massive and massless states.

\vspace{0.3cm}

\begin{center}
{\large BMS group}
\end{center}

In this case one has additional freedom since one is {\it free} to choose the topology 
for supertranslations. This is due to the fact that $\Im$ is not a Riemannian 
manifold: it is degenerate precisely in the directions along which supertranslations 
act. Having in mind Penrose description of the BMS as an {\it exact} symmetry 
acting of $\Im$,  arbitrary supertranslations 
functions describe indeed symmetry transformations along $\Im$. The standard
choice \cite{Mc1}, \cite{Mc4}
made in the literature is Hilbert or nuclear topology. The former should be associated with 
bounded systems (for which indeed the BMS group turns out to be the 
asymptotic symmetry group as originally discovered), while the latter
 with unbounded (See section 7.3 for the role 
of unbounded systems).

We first consider the Hilbert topology-i.e. we endow $N$ with the 
ordinary $L^2$
inner product on $S^2$. 
Following \cite{Mc1} , \cite{Mc2}, we remember that the supertranslations
 space can be decomposed in a translational and a supertranslational part 
$$N=A\oplus B,$$
where only $A$ is invariant under the action of $G=SL(2,\mathbb{C})$ and
$T^4=\frac{N}{B}$. Furthermore there is also this chain of isomorphisms:
$$N\sim\hat{N}\sim{N^\prime}\sim N,$$
where $\hat{N}$ is the character space and $N^\prime$ is the dual space of $N$. 
This means that given a supertranslation $\alpha$ we can
associate to it a character $\chi(\alpha)\doteq e^{if(\alpha)}$,
where the function $f(\alpha)=<\phi,\alpha>$ and where $\phi\in N$. 

The dual space can be decomposed as $N^\prime=B^0\oplus A^0$, where $B^0$ and
$A^0$ are respectively the space of all linear functionals  vanishing on $B$ and
$A$ and where only $A^0$ is G-invariant. Also the following relations are
G-invariant i.e.
\begin{equation}\label{iso} 
(N/A)^\prime \sim A^0\;\;\; N^\prime/A^0 \sim A^\prime.
\end{equation}

In view of the isomorphism between $N^\prime$ and $N$, we can expand the
supermomentum $\phi$ in spherical harmonics as
$$\phi=\sum_{l=0}^1\sum_{m=-l}^l p_{lm}Y_{lm}+\sum_{l>1}\sum_{m=-l}^l
p_{lm}Y_{lm},$$
where the first term lies in $B^0$ and the second in $A^0$.\footnote{One can interpret 
the piece belonging to $A^0$ as composed of spectrum generating 
operators, while those in $B^0$ act on the vacuum in the context of a 
holograhic description. We thank J. de Boer for the remark.}Relying on
(\ref{iso}), we can think of the coefficients $p_{lm}$ with
$l=0,1$ as the components of the Poincar\'e momentum. Thus, we call the dual
space of $N$ the supermomentum space and define a projection map:
$$\pi:N^\prime\to N^\prime/A^0,$$
assigning to each supertranslation $\phi$ a 4-vector $\pi (\phi) =(p_0,p_1,p_2,p_3)$.

At the end of the day one ends up with \cite{Mc2}

\vspace{0.3cm}

\begin{center}
\begin{tabular}{|c|c|c|} 
\hline
Little group & orbit invariant & representation label \\
\hline
$SU(2)$ & $p^2=m^2$, sgn($p_0$) & discrete spin $j$ (dim=2j+1)\\
\hline
$\Gamma$ & $p^2=m^2$, sgn($p_0$)& discrete spin $s$\\
\hline
$\Gamma$ & $p^2=0$, sgn($p_0$)& discrete spin $s$\\
\hline
$\Gamma$ & $p^2=-m^2$, & discrete spin $s$\\
\hline
$\Theta$ & $p^2=m^2$, sgn($p_0$) & discrete spin $s$\\
\hline
$\tilde{C}_n$ & $p^2=m^2$, sgn($p_0$) & finite dimensional $k$, \\ 
\hline
$\tilde{C}_n$ & $p^2=0$, sgn($p_0$) & finite dimensional $k$, \\ 
\hline
$\tilde{C}_n$ & $p^2=-m^2$, & finite dimensional $k$, \\ 
\hline
$\tilde{D}_n$ & $p^2>0$, sgn($p_0$) (for $p_0>0) $& finite dimensional $d_n$,\\ 
\hline
$\tilde{D}_n$ & $p^2<0$ & finite dimensional $d_n$,\\ 
\hline
$\tilde{T}$ & $p^2>0$, sgn($p_0$) & finite dimensional $t$. \\ 
\hline
$\tilde{O}$ & $p^2>0$, sgn($p_0$) & finite dimensional $o$. \\ 
\hline
$\tilde{I}$ & $p^2>0$, sgn($p_0$) & finite dimensional $i$. \\ 
\hline
\end{tabular}
\end{center}

\vspace{0.3cm}

\noindent Therefore the
general kets of the BMS group for massive and non massive
particles\footnote{In the BMS group the massive and the massless kets are both labelled
by discrete quantum numbers related to faithful representations of (almost the
same) compact groups whereas in the Poincar\'e case massless states are labelled by the
discrete number of the unfaithful representation of the non compact group $E(2).$} are:
\begin{equation}\label{ketBH}
\mid {\bf p}, j, s, k, d_n, t, o, i>,\;\;\;\mid {\bf p}, s, k>,
\end{equation}
where the new quantum numbers were originally 
interpreted as possible internal symmetries of 
bounded states \cite{Mc2},\cite{Mc5},\cite{komar} and the BMS group was indeed proposed to 
substitute the usual Poincar\'e group to label elementary particles 
due to the absence of non compact little groups and 
therefore of continuous spins. 

Choosing a different topology for supertranslations, however, one registers the 
appearance of non compact little groups in the BMS representations theory too 
\cite{Girardello}. We believe that precisely for this reason the hope to use 
BMS group to label elementary particles was abandoned. However another interpretation 
of these numbers has been suggested as we are going to see soon.

Consider then the ket for the nuclear (or finer) topology. First of all, recall that in this case it is
impossible to have an exaustive answer since not much is known about discrete
subgroups. Nevertheless we have \cite{Mc4} 

\vspace{0.5cm}

\begin{center}
\begin{tabular}{|c|c|c|} 
\hline
Little group & orbit invariant & representation label \\
\hline
$\Gamma$ & $p^2=m^2,0, -m^2$, sgn($p_0$)& discrete 1 dim. spin $s$\\
\hline
$SU(2)$ &$p^2=m^2$, sgn($p_0$) &discrete $2j+1$-dim. spin $j$ \\
\hline
$\Delta$ &$p^2=0$, sgn($p_0$) &finite dim. $\delta$ or $\infty$ dim.\\
\hline
$S_1$ &$p^2=0$, sgn($p_0$) &finite dim. $s_1$\\
\hline
$S_2$ &$p^2=0$, sgn($p_0$) &finite dim. $s_2$\\
\hline
$S_3$ &$p^2=0$, sgn($p_0$) &finite dim. $s_3$\\
\hline
$S_4$ &$p^2=0$, sgn($p_0$) &finite dim. $s_4$\\
\hline
$S_5$ &$p^2=0$, sgn($p_0$) &finite dim. $s_5$\\
\hline
\end{tabular}
\end{center}

\vspace{0.5cm}
We omit the study of little groups with  $m^2<0$ since in principle they have no physical relevance.  
Thus the general ket for the BMS group in the nuclear topology is:
\begin{equation}\label{Nket}
|{\bf p}, j,s,\left\{t_n\right\}>, \;\;\;
|{\bf p}, j, s,\delta, \left\{s_n\right\}> 
\end{equation}
where the first case refers to faithful representations with $m^2>0$ and the
index $\left\{t_n\right\}$ stands for all the representation numbers of finite
groups; the second ket instead refers
to the massless case and $\left\{s_n\right\}$ stands for all
 the representation
numbers of the non connected groups.

Note that it is because of the infinite dimensionality of
 supermomentum space that one has non-connected or even discrete little groups, since 
 one can have a lot of invariant vectors in this case. This is
 quite unfamiliar, since angular momenta are normally associated
  with connected groups 
 of rotations. From an experimental point of view this also renders problematic
the measurement of these ``Bondi spins'' as they are normally called. Indeed 
only in the case of the little group $SU(2)$ BMS representations contain a 
single Poincar\'e spin, otherwise they contain a mixture of Poincar\'e spins. Note 
also a curious fact: for $m^2>0 $
 all bosons with the same mass appear in the same multiplet while all fermions 
 with the same mass appear in the same multiplet though corresponding to different 
 representation.

\section{Wave equations}

In this Section we derive the covariant wave equations for the BMS group. 
As remarked before and in the following, nuclear (or even finer) 
topology is expected to be associated with unbounded systems, perhaps with 
infinite energy too. This would require in general Einstein equations in 
distributional sense and a different notion of conformal infinity. We therefore 
restrict to the Hilbert topology describing bounded systems in the bulk.

Canonical wave equations have been suggested in \cite{Mc1}, though 
physicists normally use covariant wave equations. To derive them we use the 
theorem contained in \cite{Asorey} which shows how to get irreducible covariant 
wave functions starting from (irreducible) canonical ones. The framework- based 
on fiber bundle techniques, is quite general and 
elegant. For definitions and notations see 
Appendix B, which we suggest to read before this Section. 

We are also going to use sort of diagrams in the discussion which, although not 
completely rigorous, may help to handle the formalism easier.

Consider then the following diagram 

$$\xymatrix{G=N\ltimes SL(2,\mathbb{C})
\ar[d]^\pi\ar[r]^{\;\;\;\;\;\;\;\;\;\;\Sigma} & GL(V)\ar[d]^{*} \\
N & F(N,V)\ar[l]^{\pi^\prime}}$$

The representations induced in the above way are usually referred as
\emph{covariant}. Since the bundle is topologically trivial i.e. $G=N\times
SL(2,\mathbb{C})$ we are free to choose a global section $s_0:N\to G$ and the
natural choice is $s_0(n)=(n,e)$; it is possible to see
from (\ref{gamma}) that $\gamma_0((n,k),n^\prime)=(0,k)$ which implies that
 the matrix $A(g,n^\prime)=\Sigma(\gamma(g,n^\prime))$ (see
Appendix B for definitions) does not depend on the choice of the supertranslation $n^\prime$ whereas
this fails for induced representations. The only problem for covariant representations
is that in general they are not irreducible even if $\Sigma$ is, but
as said a method \cite{Asorey} to compare covariant and induced
representations was formulated. Consider indeed the following ``diagram'':
$$\xymatrix{G\sim\hat{G}=\hat{N}\ltimes SL(2,\mathbb{C})\ar[d]^\pi\ar[r]^{\;\;\;\;\;\;\;\;\;\;\Sigma} & GL(V)\ar@{-->}[d]^{*} \\
\hat{N} & \hat{F}(\hat{N},V)\ar[l]^{\pi^\prime}},$$
where $\hat{N}$ is the character space of $N$ and $\hat{F}$ is the bundle
$\hat{N}\ltimes SL(2,\mathbb{C})\times_{SL(2,\mathbb{C})}V$. Thus we can introduce a $\hat{T}$
representation acting on the sections of $\Gamma(\hat{F})$:
$$\hat{T}(n,k)\hat{\psi}(k\chi)=(k\chi)(n)k\hat{\psi}(\chi),$$
which is a transposition of 
\begin{equation}\label{psisec}
\hat{T}(g)\hat{\psi}(g\chi)=g\hat{\psi}(\chi).
\end{equation}

Using the natural section (in the character space) $\hat{s}_0(\chi)=(\chi,e)$,
the action of the group $G$ on the function $\hat{f}:\hat{N}\to V$ is given from
(\ref{psisec}) by
\begin{equation}\label{psi4}
\left(\hat{T}_{s_0}(n,k)\hat{f}\right)(k\chi)=(k\chi)(n)\Sigma(k)\hat{f}(\chi),
\end{equation}

which we can refer to as the \emph{covariant wave equation}.
The relation between induced and covariant
representations can be made now choosing a fixed character on an orbit $\Omega$
(physically speaking going on shell)
and denoting with $\sigma$ the representation of $K\chi_0$ subdued by $\Sigma$.

The essence of the Theorem contained in \cite{Asorey} is that if $W$ is the canonical representation of $G$ in $\Gamma(F)$ induced from
$\chi_0\sigma$ than there exists an isomorphism of bundles
$\rho:F\to\hat{F}_\Omega$ such that the map
$R:\Gamma(F)\to\Gamma(\hat{F}_\Omega)$ defined by $R\psi(\chi)=\rho(\psi(\chi))$
satisfies
$$R\circ W=\hat{T}_\Omega\circ R,$$
which states the equivalence between $W$ and $\hat{T}_{\Omega}$. Notice
that with $\hat{T}_\Omega$ and with $\hat{F}_\Omega$ we simply refer to the
(on shell) restriction on the $SL(2,\mathbb{C})$-orbit. The above framework 
can be applied for a given group $G=N\ltimes H$ as follows:
\begin{enumerate}
\item identify all little groups $H_\chi\subset H$ and their orbits (labelled 
by Casimir invariants)
\item construct a representation induced from $H_\chi$ and choose a section for
the bundle $G\left((G/G_\chi),\pi, G_\chi\right)$,
\item construct the canonical wave equations for each little group.
\item construct the covariant wave equation starting from a representation of
$H$ upon the choice of a section of the fiber bundle $G(N,\pi,H)$. Write the
covariant wave equation in the dual space.
\item relate the induced and the covariant wave equations restricting the 
latter to the orbit of a little group and then defining a linear transformation $V$
acting on functions given by
$$ V=U_{\lambda}(s^{-1}(p)),$$
where $U$ is the representation of $H$ and $s$ is the section chosen at point
(2).
\item since the representation $U$ is in $H$ and it is reducible, it is
necessary to impose some constraints (i.e. projections) on the wave functions corresponding to the
reduction of the representation $U$ to one of the little groups. 
\end{enumerate}
\begin{center}
{\large Poincar\'e group}
\end{center}

We review  first how the above construction applies to the Poincar\'e group
$T^4\ltimes SL(2,\mathbb{C})$. For
further details we refer to the exsisting literature \cite{Group},
\cite{Group2} and \cite{Landsman}. As discussed before, little groups and 
orbits of the Poincar\'e group are well known and in particular for a
massive particle the orbit is the mass hyperboloid $SL(2,\mathbb{C})/SU(2)\sim\Re^3$ and a fixed
point is given by the 4-vector $(m,0,0,0)$. A representation for $SU(2)$ is the
usual $2j+1$-dimensional $D^j$ and a global section (see appendix of \cite{Mc2})
for $SU(2)$ can be chosen
remembering the (unique) polar decomposition for 
an element $g\in SL(2,\mathbb{C})$ given by
$g=\rho u$ where $\rho$ is a positive definite hermitian matrix and $u\in
SU(2)$. Thus a section $\eta$ can be written as:
\begin{equation}\label{sections}
\eta(\rho)=\eta(\rho[SU(2)])=(\rho,0).
\end{equation}
This concludes the first two points in our construction; a canonical wave
equation can be immediately written starting from (\ref{psi2}) as:
\begin{equation}\label{inducedP}
U^{m,+,j}\left(g\right)f(gp)=e^{ip\cdot
a}D^j[\rho^{-1}_{\Lambda_g\Lambda_p}\Lambda_g\rho_{\Lambda_p}]f(p),
\end{equation}  
where $g=(a,\Lambda)$, the exponential term is the character
 of $a\in T^4$, the $\cdot$ represents the Minkowskian internal product, 
$p$ is a point over the orbit, $U$ is a representation of
the little group over an Hilbert space $\mathcal{H}_j$ and $f(p)$ is a function in
$L^2(SL(2,\mathbb{C})\ltimes N/SU(2)\ltimes N)\otimes\mathcal{H}_j$. Notice also
that the representation is labelled by the $SU(2)$ quantum number $j$ and by the
indices $m,+$ that allow to select a unique point over the orbit through the
identification $m=K$ and $sgn(K)>0$.\\
The covariant wave equation can be written for functions $f(p)\in
L^2(T^4)\otimes\mathcal{H}_\lambda$ as:
\begin{equation}\label{covp}
\left(U^\lambda(g)f\right)(a)=U_\lambda(\Lambda)f(\Lambda^{-1}(a-a^\prime)),
\end{equation}
where $g=(a^\prime,\Lambda)$ and $U_\lambda$ is a representation of
$SL(2,\mathbb{C})$. In the dual space (\ref{covp}) becomes:
\begin{equation}\label{fcovp}
\left(U^\lambda(g)\hat{f}\right)(p)=e^{ip\cdot a}U_\lambda(\Lambda)\hat{f}(\Lambda^{-1}p).
\end{equation}
This concludes the fourth point of our construction; the reduction to the orbit
of the $SU(2)$ little group can be achieved requiring the mass condition for
each $\hat{f}(p)$:
$$\theta(p_0)(p^2-m^2)\hat{f}(p)=0,$$
which amounts to restrict \footnote{Here and in the following we denote with $\prime$ the 
measure restricted to the orbit.} the measure
$d^4(p)$ to $d\mu^\prime(p)=2\pi\delta(p^2-m^2)\theta(p^0)d^4 p$. This means that instead of
dealing with functions in $L^2(T^4,d^4 p)\otimes\mathcal{H}_\lambda$ we consider
elements of the Hilbert space $\mathcal{H}^{m,+,\lambda}=L^2(T^4,d\mu^\prime (p))\otimes\mathcal{H}_\lambda$.
In order to relate the canonical and the covariant wave equations, we 
introduce the operator
$$V=U_\lambda(\rho^{-1}_p),$$
where $\rho^{-1}_p=\eta(p)^{-1}$. This acts on functions as
$$(V\hat{f})(p)=U_\lambda\left(\rho^{-1}_p\right)\hat{f}(p).$$ 
Thus if we define the space $\mathcal{H}_\eta^{m,+,\lambda}$ to coincide as a
vector space with $\mathcal{H}^{m,+,\lambda}$ but equipped with the inner
product
$$(f,f^\prime)_\eta=\int
d\mu^\prime(p)\left(U_\lambda(\eta(p))^{-1}f(p),U_\lambda(\eta(p)^{-1})f^\prime(p)\right)_{\mathcal{H}_\lambda},$$
we see that the map $V$ is indeed unitary and substituting it in
(\ref{fcovp}) through $V^{-1}U^{m,+,\lambda}V=U_\eta^{m,+,\lambda}$ we
get:
\begin{equation}\label{covariantpoinc}
\left(U^{m,+,\lambda}(g)f\right)(p)=e^{i<\phi,a>}U_\lambda(\rho^{-1}_p\Lambda\rho_{\Lambda^{-1}p})f(\Lambda^{-1}p),
\end{equation}
which is coincident with the canonical wave equation even if 
$U_\lambda$ is still a representation of $SL(2,\mathbb{C})$. This means that in
general our wave function has more components that those needed and for this
reason, we have to introduce a suitable orthoprojection modding out the unwanted
components. This can be done introducing a matrix $\pi$ such that
$$\pi f(\bar{p})=f(\bar{p}),$$
where $\bar{p}$ is a fixed point over the orbit. Another way to express the projection operation is to consider the point
$f^\prime(\bar{p})=(U(\Lambda)f)(\bar{p})$ and the equation
(\ref{covariantpoinc}) in order to obtain the more familiar expression
$$\pi(p)f(p)=f(p),$$
where $p=L^{-1}_\Lambda\bar{p}$ and where $\pi(p)=D^{-1}(\Lambda)\pi D(\Lambda)$. Using the decomposition
$\Lambda=\rho_p\gamma$ that we introduced in order to construct the section for
the $SU(2)$-orbit and using the fact that $\pi$ commutes with $D(\gamma)$ with
$\gamma\in SU(2)$, then $\pi(p)=D^{-1}(\rho_p)\pi D(\rho_p)$; so we have 
that $\pi(p)$ transforms as a covariant matrix operator. 

Let us consider
explicitly the case of a massive particle with spin $\frac{1}{2}$; to 
preserve parity we consider the representation $D^{(\frac{1}{2},0)}\oplus
D^{(0,\frac{1}{2})}$, the matrix $\pi$ projecting away the unwanted components
is: 
$$\pi= diag[1,0,0,0]=\frac{1}{2}(\gamma_0+I),$$
where $\gamma_0$ is the usual Dirac's $\gamma$ matrix. Using the covariant
transformation of this matrix, we find that
$$(D^{(\frac{1}{2},0)}\oplus
D^{(0,\frac{1}{2})})^{-1}(\Lambda_p)\pi \left(D^{(\frac{1}{2},0)}\oplus
D^{(0,\frac{1}{2})}(\Lambda_p)\right)=\frac{1}{2m}(\gamma_\mu p^\mu+m),$$
where $p=L_\Lambda \bar{p}$ and where $\bar{p}=(m,0,0,0)$ is a fixed point in
the $SU(2)$-orbit. Thus the equation 
$\pi(p)f(p)=f(p)$ for $f(p)\in
L^2(\Re^3,d\mu(p))\otimes\mathcal{H_j}$ becomes:
$$(\gamma_\mu p^\mu-m)f(p)=0,$$
which is the well known Dirac equation. In a similar way we can find the 
well known equations for all $SU(2)$ spins. 

Let us briefly remark that
for massless particles the situation is more complicated since in
this case we have to deal with the non compact $E(2)$ group. The
representation of $SL(2,\mathbb{C})$ cannot be fully decomposed into
representations of $E(2)$ and thus the orthoprojection condition has to be
modified \cite{Asorey}.
\subsection{Wave equations for the BMS group}

Recalling the discussion on representations of the previous Section 
we focus on the Hilbert topology case and derive the covariant wave equations for 
the little groups $SU(2),\Gamma,\Theta$ and finite groups. 
\begin{center}
{\large The group SU(2)}
\end{center}
Clearly the first point of the construction has already been given 
by McCarthy (\cite{Mc1}, \cite{Mc2}) togheter with a partial classification of the
orbits. For massive particles a great difference arises from the Poincar\'e
group where the only orbit with $m>0$ is the one of the $SU(2)$ little group
whereas for the BMS group, apart from $SU(2)$, we need to consider more groups
as it can be seen from the tables in the previous Section.
 
Let us now address point (2). Remember that in this case the orbit is
isomorphic to $\Re^3$ and that a fixed point on the orbit is given by a constant
function $K$ over the sphere $S^2$. Thus the orbit is uniquely characterized by
the mass $m^2=K$ and the momentum 4-vector is given by $\pi(\phi)=(m,0,0,0)$
where $\phi$ is a constant supertranslation. A section for the bundle $G\left(SL(2,\mathbb{C})\ltimes
N/SU(2)\ltimes N,\pi,SU(2)\ltimes N\right)$ has been given in the previous
discussion on the Poincar\'e case through the
polar decomposition $g=\rho u$ with $u\in SU(2)$ and $\rho$ a positive
definite hermitian matrix. Thus a section $\eta$ can be written as:
\begin{equation}\label{sectionSU}
\eta[p]=\eta[\rho SU(2)]=\left(\rho,0\right).  
\end{equation} 

Point (3) is also easy to implement starting from the well known
representations of $SU(2)$ since (\ref{psi2}) becomes:
\begin{equation}\label{inducedSU}
U^{m,+,j}\left(g\right)u(gp)=e^{i<\phi,a>}D^j[\rho^{-1}_{\Lambda_g\Lambda_p}\Lambda_g\rho_{\Lambda_p}]u(p),
\end{equation}
where $g=(a,\Lambda)\in BMS$, the exponential term is the character of the 
supertranslations $a$ expressed
via the Riesz-Fischer theorem, $p$ is a point over the orbit, $U$ is a representation of
the little group over an Hilbert space $\mathcal{H}_j$ and $u(p)$ is a function in
$L^2(SL(2,\mathbb{C})\ltimes N/SU(2)\ltimes N)\otimes\mathcal{H}_j$.

After completing point (3), we can write the
general covariant wave equation. In this case we deal with
functions $f \in L^2(N)\otimes\mathcal{H}_\lambda$ ($\lambda$ is index for an
$SL(2,\mathbb{C})$ representation) i.e.
\begin{equation}\label{covariant}
\left(U^\lambda(g)f \right)(\phi)=U_\lambda(\Lambda)f (\Lambda^{-1}(\phi-\phi^\prime)),
\end{equation}
where $g=(\phi^\prime,\Lambda)$ and $U_\lambda$ is a representation of
$SL(2,\mathbb{C})$. We go to the dual space of supertranslations 
using the well defined character $\chi(\alpha)$ and 
functional integration restricting to the $SU(2)$ orbit to get 
for $\hat{f}(a)\in L^2(\hat{N})\otimes\mathcal{H}_\lambda$.
\begin{equation}\label{covariant2}
\left(U^\lambda(g)\hat{f}\right)(a)=e^{i<\phi^\prime,a>}U_\lambda(\Lambda)\hat{f}(\Lambda^{-1}a),
\end{equation}
Thus in the end we deal with functions in the Hilbert space
$\mathcal{\hat{H}}^{m,+,\lambda}=L^2(N,d\mu^\prime)\otimes\mathcal{H}_\lambda.$
The linear operator relating the induced and the covariant wave equations is
given for a generic point in the orbit $p=(\phi,\Lambda)$ by
$$V=U_\lambda(\rho^{-1}_p),$$
where $\rho^{-1}_p=\eta(p)^{-1}$. This acts on functions as
$$(V\hat{f})(p)=U_\lambda\left(\rho^{-1}_p\right)\hat{f}(p).$$ 
Thus if we define the space $\mathcal{H}_\eta^{m,+,\lambda}$ to coincide as a
vector space with $\mathcal{H}^{m,+,\lambda}$ but provided with the inner
product 
$$(f,f^\prime)_\eta=\int_{orbit}
d\mu^\prime(p)\left(U_\lambda(\eta(p))^{-1}f(p),U_\lambda(\eta(p)^{-1})f^\prime 
(p)\right)_{\mathcal{H}_\lambda},$$
we see that the map $V$ is indeed unitary and substituting it in
(\ref{covariant2}) through $V^{-1}U^{m,+,\lambda}V=U_\eta^{m,+,\lambda}$ we
get:
\begin{equation}\label{covariant3}
\left(U^{m,+,\lambda}(g)f\right)(p)=e^{i<\phi,a>}
U_\lambda(\rho^{-1}_p\Lambda\rho_{\Lambda^{-1}p})f(\Lambda^{-1}p),
\end{equation}
which is coincident with the canonical wave equation though 
$U_\lambda$ is still a representation of $SL(2,\mathbb{C})$.

The last point of our construction comes into play since we need to 
impose further constraints on the wave equation in order to mod out the unwanted
components arising from the fact that $U_\lambda$ is a representation of
$SL(2,\mathbb{C})$ that has to be restricted to an $SU(2)$ one. This reduction can be expressed
using a matrix $\pi$ and, for the BMS group, the discussion for $SU(2)$ is
exactly the same as in Poincar\'e; so if we choose a fixed point $\bar{\phi}$ in the orbit, the constraints we have to impose becomes:
\begin{equation}
\pi[SU(2)]f(\bar{\phi})=f(\bar{\phi}).
\end{equation}
Remembering that a representation $D^{(j_1,j_2)}\in SL(2,\mathbb{C})$ can be
decomposed into SU(2) representations through
$D^{(j_1,j_2)}=\bigoplus\limits_{j=\mid j_1-j_2\mid}^{j_1+j_2}D^j$, the matrix
$\pi$ simply selects the desired value of $j$ from the above decomposition. If,
as an example, we consider the value $j=1$ in the representation
$D^{(\frac{1}{2},\frac{1}{2})}=\bigoplus\limits_{j=0}^1 D^j$, the matrix $\pi$ is:
$$\pi[SU(2)]=diag[0,1,1,1].$$

\begin{center}
{\large The group $\Gamma$}
\end{center}

Consider now $\Gamma$, namely the double cover of $SO(2)$. An element of this group is given by a
diagonal complex matrix:
$$g=\left[\begin{array}{cc}
e^{\frac{i}{2}t} &0\\
0& e^{-\frac{i}{2}t}
\end{array}\right].$$ Again in \cite{Mc2} the orbits for this little group were 
studied and it was shown that both the squared
mass and the sign of the energy are good labels. A fixed point on this orbit is
given by a supertranslation depending only on the modulus of the complex
coordinate over $S^2$ i.e. $\psi=\psi(\mid z\mid)$ (or equivalently in real
coordinates this means that $\psi(\theta,\varphi)$ does not depends on $\phi$).Thus
in this case the projection over the four-momentum for the fixed point is
$$\pi(\psi(\theta,\varphi))=(p_0,0,0,p_3).$$

Notice also that we choose in the orbit of $\Gamma$ those functions not in
the orbit of $SU(2)$-i.e. they cannot be transformed into a constant function;
this means that any point in the orbit has the form $\psi=\Gamma\psi(\mid
z\mid)$. In \cite{Mc1} also a section for the bundle
$G(SL(2,\mathbb{C})/\Gamma,\pi,\Gamma)$ has been given: any element $g\in SL(2,\mathbb{C})$ can be decomposed as explained
before as $g=\rho u$ with $u\in SU(2)$. The element $u$ can further be
decomposed as $u=\gamma_\phi\sigma_\theta\gamma_\psi$ which implies
$g=\tau\gamma_\psi$ where now $\gamma_\psi\in\Gamma$. Thus a section can be
written as:
$$\eta(\tau[\Gamma])=(0,\tau)\in BMS.$$
This completes the first and the second point of the construction. The canonical
wave equation can be written directly from the one-dimensional representation
for $\Gamma$ acting on an element $\gamma$ as a multiplication in one complex
dimension:
$$D^s(\gamma)=e^{ist}.$$
This leads to
\begin{equation}\label{inducedg}
U^{m,+,s,p_0}(g)f(gp)=e^{i<\phi,a>}D^s(\tau^{-1}_{\Lambda_g\Lambda_p}\Lambda_g\tau_{\Lambda_p})f(p),
\end{equation}
where $g=(a,\Lambda)\in BMS$, $f(p)$ is a square integrable function over the orbit
$SL(2,\mathbb{C})/\Gamma\sim \Re^3\times S^2$ and the indices on the 
left hand side label the orbit. 

Notice that the difference from $SU(2)$ is that these
indices are not uniquely determining the orbit since in that case the fixed
point was given by the constant supertranslation $K=m$ and for this reason the
value of the mass, and the sign of $p_0$ were fixing in a unique way the point over the
orbit. On the other hand here the labels $m,+,p_0$ are fixing only a class of
points with the same 4-momentum i.e. $p^\mu=(p_0,0,0,\sqrt{m^2+p^2_3})$ whereas
the supertranslation associated is not unique since in real coordinates it has
the form:
$$\phi(\theta)=p_0+p_3cos\theta+h(\theta),$$
for all possible $h(\theta)$ $\Gamma$-invariants.

It also worth stressing that if we choose $p_0=m$ and  $p_3=0$ (i.e. the
equivalent of the rest frame), the associated supertranslation is not constant
but it has the form $\phi(\theta)=K+h^\prime(\theta)$ with $h^\prime\neq 0$
since otherwise the point would belong to the $SU(2)$ orbit which is
impossible. This can be seen as the
impossibility to define ordinary angular momentum . More generally this is related to
the statement \cite{Mc8} that the representations of BMS different from $SU(2)$
cannot be uniquely reduced to Poincar\'e ones but each of them decomposes into
many differents Poincar\'e spins as recalled before.

Point (4) of our construction goes along similar lines 
just by reducing the functional integration to the corresponding orbit. 
Thus we only need to define the linear map for a generic point over
the orbit $p=(a,\Lambda)$ 
$$V=U(\rho_p^{-1}),$$
where $\rho^{-1}(p)$ is the inverse of the section $\eta(p)$.
As in the case of $SU(2)$ this linear operator switches from the space
$\mathcal{H}^{m,+,p_0,s}=L^2(N,d\mu^\prime)\otimes\mathcal{H}_s$ to the space 
$\mathcal{H}^{m,+,p_0,s}_\eta$ which coincides as a vector space but the internal product is:
$$(\psi,\phi)_\eta=\int\limits_{\Re^3\times S^2}
d\mu^\prime(p)\left(U_\lambda(\eta(p)^{-1})\psi(p),U_\lambda(\eta(p)^{-1})\phi(p)\right)_{\mathcal{H}_s}.$$
As before substituting the map $V$ into (\ref{covariant2}), we get:
\begin{equation}\label{covariant5}
\left(U^{m,+,p_0,\lambda}(g)f \right)(p)=e^{i<\phi,a>}
U_\lambda(\tau^{-1}_p\Lambda\tau_{\Lambda^{-1}p})f(\Lambda^{-1}p),
\end{equation}
which is coincident with (\ref{inducedg}) except that the representation needs to
be reduced from $SL(2,\mathbb{C})$ to $\Gamma$. Since
 all little groups in the Hilbert topology are compact, we know that $U$ is always
completely reducible and that the desired irreducible component can be 
picked out.

As before we start from a representation $D^{(0,j)}\oplus D^{(j,0)}$ to make 
sense of parity and we impose restriction. Since the decomposition
of $D^j(SU(2))$ into $\Gamma$ is well known and has the form:
$$D^l(\Gamma)=\bigoplus\limits_{m=-l}^l U^m(\Gamma),$$
we can first project from $SL(2,\mathbb{C})$ into a representation of $SU(2)$
and using the above decomposition, select a particular value of s. The
procedure is basically:
$$D^{(j_1,j_2)}(g)=\bigoplus\limits_{j=\mid
j_1-j_2\mid}^{j_1+j_2}D^j(g)=\bigoplus\limits_{j=\mid
j_1-j_2\mid}^{j_1+j_2}\bigoplus\limits_{s=-j}^jD^s(g).$$
The subsidiary condition is:
$$\pi(\Gamma)f(\bar{\phi})=f(\bar{\phi}),$$
where $\bar{\phi}$ is a fixed point over the orbit (i.e.
$\bar{\phi}=p_0+p_3\cos\theta+h(\theta)$) and where $\pi(\Gamma)$ extracts 
from the above decomposition the desired $s$
component. As an example in the case of $s=1$ from the representation
$D^{(\frac{1}{2},\frac{1}{2})}$ we know that:
$$D^{(\frac{1}{2},\frac{1}{2})}=\bigoplus\limits_{j=0}^1D^j(g)=\bigoplus\limits_{j=0}^1\bigoplus\limits_{s=-j}^jD^s(g),$$
so that 
$$\pi= diag[0,0,0,1] $$
\begin{center}
{\large The group $\Theta$}
\end{center}

The third group we examine is the compact, non-connected
little group $\Theta=\Gamma R_2$ where $R_2$ is the set (not the group) given by
the matrix $I$ and $J=\left[\begin{array}{cc}
0 & 1\\
-1 & 0\end{array}\right]$. The orbit of this group is given by the points in the
orbit of $\Gamma$ which are also invariant under $R_2$. This last condition is
equivalent to require for the supertranslation $\psi(\theta)=\psi(-\theta)$.
Upon projection over the dual space this implies that
$\pi(\psi)=p_0+p_3cos\theta=p_0$. Thus a fixed point under the action of $\Theta$ is
an element of the orbit of $\Gamma$ with the form
$$\psi(\theta)=p_0+h(\theta),$$
with clearly $h(\theta)\neq 0$ since otherwise the supertranslation would be in
the $SU(2)$ orbit.

Another important remark is that the orbit
$SL(2,\mathbb{C})/\Theta$ is $\Re^3\times P^2$ which is the same orbit of the
group $\Gamma$ plus the antipodal identification of the points over the sphere
due to $R_2$. In order to choose a section we remember that every $g\in
SL(2,\mathbb{C})$ can be decomposed as $g=\tau\gamma$ with $\gamma\in\Gamma$; a
point in $SL(2,\mathbb{C})/\Gamma$ is thus identified with the value of $\tau$.
In our case a global section (see \cite{Mc2}) can be given noticing that
every matrix $\sigma_\theta=\left[\begin{array}{cc}
\cos\frac{\theta}{2} & i\sin\frac{\theta}{2}\\
i\sin\frac{\theta}{2} & \cos\frac{\theta}{2}\end{array}\right]$ can be decomposed as
$\sigma_\theta=\sigma_{\theta^\prime}q$ with $0<\theta^\prime<\frac{\pi}{2}$,
$q=\gamma_{\frac{\pi}{2}}r\gamma_{-\frac{\pi}{2}}$ and $r\in R_2$. Since any
element $g$ of $SL(2,\mathbb{C})$ can be written as $g=\rho
u=\rho\gamma_\Phi\sigma_\theta\gamma_\psi$, we can plug in the above
decomposition 
$$g=\rho\gamma_\Phi\sigma_{\theta^\prime}q\gamma_\psi=\beta q^\prime.$$

Thus we can see that a section $\omega:G/\Theta\to G$ is given by
$\omega(g\Theta)=\omega(\lambda q^\prime\Theta)=\omega(\lambda\Theta)=\lambda$.
From this we can easily write a section 
$SL(2,\mathbb{C})\ltimes N/\Theta\ltimes N\to SL(2,\mathbb{C})\ltimes N$ as:
$$\eta(\beta)=(\beta,0).$$
This concludes point (1) and (2) of the construction. Consider now $\Theta$ 
representations: the 1-dimensional one when $s=0$ (s is the index for the
representation of $\Gamma$) which is $U(\gamma)=1$ and $U(J)=-1$ whereas for
integer $s$ we have:
$$U(\gamma_\phi)=\left[\begin{array}{cc}
e^{\frac{i}{2}s\phi} & 0\\
0 & e^{-\frac{i}{2}s\phi} 
\end{array}\right],\;\;\; U(J)=\left[\begin{array}{cc}
0 & (-)^s\\
1 & 0\end{array}\right].$$
Recalling that the for the orbit of $\Theta$ we can apply the same
considerations and the same labels as for $\Gamma$ except that in our case
$p_3=0$ (and $p_0=m$), we can write 
\begin{equation}\label{inducedt}
U^{m,+,s}(\gamma_\theta)(g)f(gp)=e^{i<\phi,a>}D^s[\beta^{-1}_{\Lambda_g\Lambda_p}\Lambda_g\beta_{\Lambda_p}]f(p),
\end{equation} 
where here we denote with $D^s$ the representation of $\Theta$ over an Hilbert
space $\mathcal{H}_s$ and thus the function $f(p)$ is in
$\mathcal{H}^{m,+,s}=L^2(SL(2,\mathbb{C})\ltimes N/\Theta\ltimes
N)\otimes\mathcal{H}_s$. We proceed than as in the previous cases. We introduce the operator
$V=U_s(\eta(\beta(p)^{-1}))$ that sends the Hilbert space $\mathcal{H}^{m,+,s}$ to
$\mathcal{H}^{m,+,s}_\eta$ which is coincident as a vector space to the first
but it is endowed with the internal product:
$$(f,f^\prime)_\eta=\int\limits_{\Re^3\times P^2}
d\mu^\prime(p)\left(U_\lambda(\eta(p))^{-1}f(p),U_\lambda(\eta(p)^{-1})f^\prime(p)\right)_{\mathcal{H}}.$$
As usual substituting the map $V$ into (\ref{covariant2}) we get:
\begin{equation}
\left(U^{m,+,\lambda}(g)f\right)(p)=e^{i<\phi,a>}
U_\lambda(\beta^{-1}_p\Lambda\beta_{\Lambda^{-1}p})f(\Lambda^{-1}_p),
\end{equation}
where $f(p)=(V\hat{f})(p)$ and $U_\lambda$ is the restriction of
$U_\lambda$ from $SL(2,\mathbb{C})$ to $\Theta$. We now reduce
$U_\lambda$ and this is indeed possible since $\Theta$ is compact. 
This can be achieved as for the $\Gamma$
case using the character formula (see \cite{Mc8}) and every single $s$ appears
exactly once in the decomposition of $SU(2)$. We proceed then as in 
the $\Gamma$ situation with the exception that now the
projection equation $\pi[\Theta]f(\bar{\phi})=f(\bar{\phi})$ is applied to the
supertranslations which are a fixed point over the orbit of the $\Theta$ group. The form of
the matrix will indeed be the same.
\begin{center}
{\large Finite groups}
\end{center}

Only finite dimensional groups have to be considered in the massive
case. We shall address only the (double cover of the) cyclic group $C_n$ as an example since all the
others are similar cases. 

The group
$\tilde{C}_n$ is given by the diagonal matrices:
$$c_n=\left[\begin{array}{cc}
e^{\frac{\pi ik}{n}} & 0\\
0 & e^{\frac{-\pi ik}{n}}\end{array}\right],$$
where $1\leq k\leq 2n$. The orbit for this group is constructed from a fix point
which is given by those supertranslations satisfying the periodic condition
$\psi(\theta,\varphi)=\psi(\theta,\varphi+\frac{2\pi}{n})$. Thus, a part from the case
$n=1$ which is trivial, the above condition tells us that the function
$\psi(\theta,\varphi)=p_0+p_3\cos\theta+k(\theta,\varphi)$ where $k(\theta,\varphi)$ is a
pure supertranslation. Furthermore we can assign a global section to this orbit
noticing that any element of $\Gamma$ can be decomposed as:
$$\left[\begin{array}{cc}
e^{i\frac{t}{2}} &0\\
0& e^{-i\frac{t}{2}}\end{array}\right]=\left[\begin{array}{cc}
e^{i\frac{t^\prime}{2}} &0\\
0& e^{-i\frac{t^\prime}{2}}\end{array}\right]\left[\begin{array}{cc}
e^{i\frac{\pi k}{n}} &0\\
0& e^{-i\frac{\pi k}{n}}\end{array}\right],$$
where $0<t^\prime<\frac{\pi}{n}$. Since any element $g$ of
$SL(2,\mathbb{C})$ can uniquely be written as $g=\tau\gamma_\theta$, we can plug
in the above decomposition writing $g=\tau\gamma_{\theta^\prime}c_k$. Thus
$\omega(gC_n)=\omega(\tau\gamma_{\theta^\prime}c_kC_n)=\omega(\alpha C_n)=\alpha$.
This concludes both points (1) and (2) of our construction. The representation is
simply one dimensional and gives $D^k(g)=e^{\frac{ik\pi}{n}}$ with $1\leq k\leq
2n$. Thus the canonical wave equation is simply:
\begin{equation}\label{inducedc}
U^{m,+,p_0,k}(g)f(gp)=e^{i<\phi,a>}D^k[\alpha^{-1}_{\Lambda_g\Lambda_p}\Lambda_g\alpha_{\Lambda_p}]f(p),
\end{equation}
where as usual $g=(a,\Lambda)$, $p$ is a point over the orbit and $f(p)\in
L^2(SL(2,\mathbb{C}\ltimes N/C_n\ltimes N)\otimes\mathcal{H}_k$. 
The operator relating the two description is as usual $V=U_\lambda(\eta_p^{-1})$ that
acting on functions over $\mathcal{H}^{m,+,p_0,\lambda}=L^2(N,
d\mu^\prime)\otimes\mathcal{H}_\lambda$ as:
$$(V\hat{f})(p)=U_\lambda\left(\eta^{-1}(p)\right)\hat{f}(p).$$
Thus V sends the Hilbert space $\mathcal{H}^{m,+,p_0,s}$ into
$\mathcal{H}^{m,+,p_0,s}_\eta$ which are coincident as vector spaces but the
latter is endowed with the inner product:
$$(f,f^\prime)_\eta=\int\limits_{\Re^3\times\frac{P^3}{C_n}}
d\mu^\prime\left(U_\lambda(\eta(p)^{-1})f(p),U_\lambda(\eta(p)^{-1})f^\prime(p)\right)_{\mathcal{H}_\lambda}.$$
Substituting the map $V$ into (\ref{covariant3}), we get 
\begin{equation}\label{covariant10}
\left(U^{m,+,p_0,\lambda}f\right)(p)=e^{i<\phi,a>}
U_\lambda(\alpha^{-1}_p\Lambda\alpha_{\Lambda^{-1}_p})f(\Lambda^{-1}p),
\end{equation} 
which is as usual coincident with (\ref{inducedc}) except for the fact that we
need some further constraints to select the rep we like. In this case we can start from a representation of
$SL(2,\mathbb{C})$ and progressively reduce it first to $SU(2)$ then to $\Gamma$
and in the end to $C_n$. Following \cite{Mc8} we only need to consider the
reduction of a representation $D^s$ of $\Gamma$. Since on an element $c_n\in
C_n$ $D^s$ acts giving $e^{\frac{i\pi s}{n}}$ whereas the representation of
$D^k$ of $C_n$ gives $e^{\frac{i\pi k}{n}}$; thus the representation of $C_n$
appears only one or no times in $D^s$. This condition is expressed by the
equation $s=\frac{k}{2}(mod\; n)$. From this we can see that
$$D^{(j_1,j_2)}(g)=\bigoplus\limits_{j=\mid j_1 - j_2\mid}^{j_1
+j_2}\bigoplus\limits_{s=-j}^j\delta_{s,\frac{k}{2}(mod\; n)}D^k(g).$$
We can easily  now extract the orthoprojection matrix $\pi$ and write
the additional conditions
$$\pi[C_n]f(\bar{\phi})=f(\bar{\phi}),$$
where $\bar{\phi}$ is a fixed point over the orbit of the $C_n$ group.
As an example let us consider the case $k=0$ for $C_2$ in the representation
$D^{(\frac{1}{2},\frac{1}{2})}$. We find from the decomposition that the
$U^0(C_2)$ can appear only when the equation $0=\frac{p}{2}(mod 2)$ holds; since $p$
ranges only from $-1$ to $1$, this implies that the desired representation appears
only once in each $D^0(\Gamma)$.Thus 
$$\pi[C_2]=diag[1,0,1,0].$$

This concludes our analysis for the massless case and also for the massive
case, since all other discrete groups in the Hilbert topology are acting like the
cyclic. 
\subsection{Generalizations of the BMS group and possible extensions of the results}

A natural question arising both from the representation theory and wave
equations concerns the origin and meaning
 of discrete little groups. A suggestion comes from \cite{Mc7} where McCarthy studies all
possible generalizations  (42 at the end) of the BMS group. 
Among them one finds $N(S)\ltimes L_+$ where $L_+$ is the usual (connected component of the
homogeneous) Lorentz group and $N(S)$ is
the set of $C^\infty$ scalar functions (``supertranslations'' this time 
defined on a hyperboloyd and depending on three angular coordinates) from
$S=\left\{x\in\mathbb{R}^{3,1}\;\mid\;x\cdot x<0\right\}$ to $\mathbb{R}$. This
group is isomorphic to the {\it Spi group} identified by Ashtekar and Hansen
\cite{hansen} as the asymptotic symmetry group of spatial infinity $i_0$ in
asymptotically flat space-times. The study of the representations for this group
as well as for all others BMS-type groups can be carried on exactly along the 
lines of the
original BMS. Thus we have still a freedom on the choice of topology for the
``supertranslation'' subgroup and wave equations can be in principle derived
exactly in the same way as we did in the previous sections. Furthermore this
suggests that a candidate field theory living on $i_0$ with fields carrying
representations of $Spi$, should display, as well as the
theory on $\Im^\pm$, a high degree of non locality this time with even 
more degeneracy since three instead of two angular coordinates define 
supertranslations.

Finally McCarthy also identified the euclidean BMS and the
complexification of the BMS group\footnote{The euclidean BMS is the semidirect product
of $N(\mathbb{R}^4-\left\{0\right\})$ with $SO(4)$ whereas the complexification
of BMS is given by the semidirect product of the complexified Lorentz group
$\mathbb{C}SO(3,1)$ with the space of scalar functions from
$\mathcal{N}=\left\{x\in\mathbb{C}^4\;\mid\;x\cdot x=0,\;\;x\neq 0\right\}$ to
$\mathbb{C}$.}; the study of representations for these groups endowed
with Hilbert topology gives rise to discrete groups and it has been suggested to
relate them to the parametrization of the gravitational instantons 
moduli space. 

\section{Implications for the holographic mapping}

\subsection{Identifying boundary degrees of freedom}

As we have seen in the previous analysis wave functions appearing in covariant and 
canonical wave equations are functions of the supertranslations or in dual terms of 
supermomenta. We have therefore a huge degree on non locality, the fields
 depending 
on an enormous (actually infinite) number of parameters entering
 into the expansions of supertranslations (supermomenta)
 on the 2-spheres.  
 
In addition their fluctuations are supposed to spread out on
 a degenerate manifold at null infinity. Recall also that 
supertranslations act along the $u$ direction and are punctual transformations, with $u$ playing 
the role of an affine parameter. Because of the difficulties in defining a theory 
on a degenerate manifold we find therefore more natural to place these fields and the 
putative boundary dynamics on the so 
called cone space \cite{bramson}, the space of smooth cross sections on $\Im$. The BMS group is 
thus interpreted as the group of mappings of the cone space onto itself.

We can give a pictorial description as follows: fix a two sphere
 section on $\Im$ and associate 
with this a point in the cone space calling this the origin.
 Any other point in the cone space will correspond to another 
two sphere section on $\Im$ and can be obtained by moving an affine distance $u=\alpha
(\theta,\phi)$ along the original $\Im$. In this way points on cone space are mapped one to 
one to cross sections on $\Im$. 

Therefore the holographic data ought to be encoded\footnote{See 
\cite{Jan} for a similar considerations despite differences in the choice 
of screens.} on the set of 2-spheres and these in 
turn are mapped to points in cone space; we think then 
the candidate holographic description living in an 
abstract space, implying at the end of the day that 
holography should be simply an equivalence of bulk amplitudes with those derived 
from the boundary theory.\footnote{The situation can be compared to the 
BMN \cite{BMN} limit of AdS/CFT, where the boundary of a pp-wave background
 is a null one 
dimensional line and geometrical interpretation seems difficult (and may be lacking). Again 
holography seems to be thought as an equivalence between bulk/boundary amplitudes, the latter 
may be living in some smaller CFT.}

An interesting consequence of working in the cone space is that one has in principle a 
way to define a length and therefore separate, in some sense, the spheres $S^2$ along null infinity. 
One can actually choose coordinates on the cone space: they will be 
the coefficients entering in the 
expansion of supertranslations in spherical harmonics. One can then show that
there exsists 
an affine structure (infinite dimensional) on the cone space and eventually define a 
length for vectors in the cone space \cite{bramson}
\be \label{lunghezza}
L^2 = [\int d\Omega (\alpha(\theta,\phi))^{-2}]^{-1}
\ee
This should allow to define \cite{noi} a sort of ``cutoff'', a concept otherwise absent on the 
original degenerate $\Im$. In the case of AdS/CFT the dual theory is a CFT with 
no fundamental scale. There, however, one uses the fact that AdS is basically 
a ``cylinder'' to end up with the correct counting from both sides \cite{susskind}.

This goes along the direction suggested
 by Bousso in \cite{changing}. 
 Actually apart from the special AdS case where
 one can show that, moving the boundary-screen 
to infinity, the boundary theory is indeed dual since it contains no more than one degree 
of freedom per Planck area \cite{susskind}, the ``dual theory''
 approach should not work in our case and one 
should expect theories with a {\it changing} number of degrees of freedom in the case of 
null boundaries. Degrees of freedom 
should appear and disappear continuously\footnote{We thank G.'t Hooft for
stressing this point also in cosmological context}. The
 dependence of (\ref{lunghezza}) on the coefficients 
tells us that a possible cut off length can change according to the 
number of coefficients we switch on-off. In turn we have seen that
 again $\alpha(\theta ,
\phi)$ enters in the changing of asymptotic shears and we have related the possibility 
of more/less bulk production according to the vanishing of asymptotic shear. 

Interestingly, more/less bulk entropy will have at the end of the day effect on
the way one defines lengths on the 
cone space.

\subsection{Similarities with 't Hooft S-matrix Ansatz for black holes}
The final picture one gets is quite similar to the scenario proposed by 't Hooft
\cite{gerard} in the context of black holes. In this case we have sort of holographic fields 
living on the horizon of the black hole. Time reversal symmetry is 
required and therefore one has operators living on the future and past horizons. 

The description is given in first quantized set up and the degree of non 
locality is eventually expressed in the operator algebra at the horizon
\be
[u(\Omega),v(\Omega')]=i f(\Omega - \Omega')
\ee
\noindent where $u(\Omega),v(\Omega)$ are the holographic fields living on the 
future/past horizon depending on the angular coordinates $\Omega$ 
and $f(\Omega,\Omega')$ is the Green function of the Laplacian operator 
acting on the angular horizon coordinates. Clearly the 
algebra is non local.

In this case too the angular coordinates are at the end of the day responsible 
for the counting of the degrees of freedom, even if 't Hooft S-matrix Ansatz
is derived in a sort of eikonal limit assuming therefore a resolution bigger than 
Planck scale in the angular coordinates.

The Green function $f(\Omega,\Omega')$ tells 
us how to move on the granular structure living on the horizon and it is similar 
to our supertranslations generated indeed by $P_{lm}=Y_{lm}(\Omega) \partial_u$. 
The holographic information is therefore spread out in both cases on angular 
coordinates.

Going then to a second quantized description of 't Hooft formalism, fields are 
expected to be functional of $u(\Omega)$ and $v(\Omega)$, pretty much in the same 
way of our case. A proposal of 't Hooft (for the 2+1 dimensional case) preserving 
covariance is indeed
\be
\phi \sim \sum_{orderings} \int d\Omega \left(
\delta(x^1(\Omega)-x) \delta(x^2(\Omega)-y) \delta(x^0(\Omega)-t) \right)
\ee
where fields are functionals of coordinates in turn depending on the angles. 
Of course the horizon itself is a very special null surface 
and the set up is different, since the whole 't Hooft picture is dynamically generated in a holographic reduction 
taking place in a sort of WKB limit. At the end of the day, however, angular 
coordinates and their resolutions are the basis for the book-keeping of states.

The horizon itself is a sort of computer storing-transmitting information. One looses 
in a sense the notion of time evolution. This suggests that also in our case (recall 
that $u$ acts via point transformations) the fields we have constructed 
are not quite 
required to evolve but are independent data living on the
2-spheres. What generates the dynamics should be a S-matrix in the spirit 
of 't Hooft. In this sense, the states are indeed holographic, since contain all 
bulk information. 
\subsection{Remarks on the construction of a S-matrix}
Therefore, in analogy with 't Hooft scenario, we 
could imagine \cite{noi} a S-matrix with in and out states
living on the respective cone spaces corresponding to  $\Im^+$ and
$\Im^-$ with fields carrying BMS labels. Of course the big task is to 
explicitly construct such a mapping.\footnote{One should also may be take into account 
the role of spatial infinity in gluing the past and the future null 
infinities.}

However, the motivation for a S-matrix is also due to the fact that in the asymptotically 
flat case we have problems with massive states which can change the 
geometry at infinity. In the case of AdS/CFT correspondence, on the other 
hand, it is true that one has a sort of box with walls at infinity so that
quantization of modes is similar to fields in a cavity. But via the Kaluza-Klein 
mechanism one generates a confining potential for massive modes. Note also 
that already for massless fields the scattering problem in the physical spacetime
 can be translated in a 
characteristic initial value problem \cite{frolov} at null infinity in the unphysical 
spacetime.\footnote{\cite{frolov} contains a derivation of the Hawking 
effect and an interesting discussion on  the BMS group. The main point is that
one can have in any case an unambiguous definition of positive/negative frequencies and this 
is what matters for the Hawking particle production. Recall however that there ones refers 
to fields  and their asymptotics in the bulk, not to fields carrying 
BMS representations as the ones we have derived.}

The way in which one should proceed in quantizing has to be
different from the usual one, since one does not need choices of
polarizations
to kill unwanted phase space ``volume''. This has already been pointed out
in \cite{Mc9} discussing BMS representations and therefore is
automatically
induced to fields carrying BMS representations we have constructed.
                                                                                
The choice of Hilbert topology, as already remarked, should be associated
with
bounded sources in the bulk, while the nuclear one should correspond to
unbounded
systems. In order to accommodate the unbounded systems which ought to
correspond to scattering states, one should however define a proper notion
of
conformal infinity for unbounded states and this seems a difficult
problem. Note however that to have a unitary S-matrix in a candidate
holographic
theory one {\it must} include unbounded
states into the Hilbert space, otherwise
asymptotic completeness \cite{simon} is violated.
                                                                                
In \cite{Mc4} it was also suggested to take a finer topology than the
nuclear one because of the freedom in the
topology choice for the supertranslations. It was proposed to use real
analytitc functions enlarging therefore supermomenta to real
hyperfunctions on the sphere. Interestingly quantum field theories in
which
fields are smeared by hyperfunctions show a non local behaviour and the
density of states can have a non polynomial growth.
This
might in principle allow to recover bulk locality\footnote{See \cite{Giddings}
for similar considerations even if from a different point of view.} although one
should consider hyperfunctional
solutions to the Einstein's equations. Moreover, if one assumes that the high
energy behaviour of the density of states in the bulk is dominated by black
holes, the exponential growth of states which suggests an intrinsic degree of
non locality might be explained by working with hyperfunctions. This is again in
sharp contrast with the asymptotically AdS case where the black hole density of
states grows essentially like the entropy of a CFT \cite{tom}.

\section{Concluding remarks}

Previous considerations show that holography in asymptotically flat 
spacetimes is in principle quite different from the usual paradigm in which 
we are expected to project information following bulk propagation. 

We have seen the relation between bulk entropy production and 
boundary symmetries and the nature of the fields carrying BMS representations.
 We have then suggested a possible way to 
encode bulk data. An equally interesting possibility to construct a model
which realizes
the BMS is to start from AdS/CFT and isolate a
"flat space region" inside AdS. It follows then that the sector of the
dual theory that appears to describe this would be the deep infrared,
namely the  uniform modes, or a matrix model in the case of the SYM dual
to AdS.\footnote{The appearance of a Matrix model in the BFSS dual to
M-theory also suggests a matrix model dual to flat space. We thank
 the referee for this remark.}

This does not mean that holography does not work but it seems likely it has 
to be thought in more general terms
 as an identity between theories living in different dimensions and this point of view 
has already been adopted in various contexts. After all the picture of light rays propagation 
is classical and (near) BPS D-branes and their charges ``prefer'' AdS in the large 
N double scaling limit corresponding to classical supergravity.

One might 
of course think as often done about
topological phases of gravity, reabsorbing in such a way the infinite number of 
parameter due to non renormabizability in a topological field and pushing then 
in a natural way degrees of freedom to the boundary. 
However gauge non invariant excitations seems to be the relevant ones in the usual picture 
of physics we have and the phase transition from the topological phase seems 
quite difficult to explain. Formulating a holographic description in 
asymptotically flat spacetimes remains then a challenging and very open problem.

\begin{center}
{\bf Acknowledgments}
\end{center}
We thank G. 't Hooft, M. Trigiante and M. Carfora for discussions and useful
comments. C.D. thanks R. Loll and Spinoza Institute for hospitality during the
realization of this work. The work of G.A. is supported in part by the European Community's Human 
Potential Programme under contract HPRN-CT-2000-00131 Quantum Spacetime.

\appendix
\section{Induced representations for semidirect product groups}
We review in the following the theory of induced representations for 
semidirect product groups. Notations and conventions are those of \cite{Simms}. 

Choose a locally compact group $G$ and a closed subgroup $K\subset G$
whose unitary representations $\sigma: K\to U(M)$ on a Hilbert space $M$ are
known. On the topological product $G\times V$, define an
equivalence relation
$$(gk,v)\sim (g,\sigma(k)v)\;\;\;\forall k\in K,$$
and the natural map
\begin{equation}\label{hilb}
\pi:G\times_K V\to G/K,
\end{equation}
assigning to equivalence classes $[g,k]$ the element $gK$ on the coset $G/K$.
The structure given in (\ref{hilb}) is clearly the one of a fiber bundle where the generic fiber over a base point 
($\pi^{-1}(gK)=\left\{[g,v]\right\}$) uniquely determines the element $v\in M$.
This bijection gives to $\pi^{-1}(gK)$ the structure of a Hilbert
space\footnote{A Hilbert bundle is defined as $\pi:X\to Y$ where both $X,Y$
are topological spaces and a structure of a Hilbert space is given to
$\pi^{-1}(p)$ for each $p\in Y$.}.

One can also introduce then a {\it Hilbert G-bundle} which is a Hilbert
bundle as before with the action of a group $G$ on both $X,Y$ such
that the map 
$$\alpha_g:x\longrightarrow gx,\;\;\;\beta_g:y\longrightarrow gy$$ is a Hilbert
bundle automorphism for each group element.\\
Choose then a unique invariant measure class on the space $G/K$ defined as
above-i.e. for any $g\in G$, for any Borel set $E$ and for a given measure $\mu$ also
$\mu_g(E)=\mu(g^{-1}E)$ is a measure in the same class. It can be
extended to any G-Hilbert bundle $\zeta=(\pi:X\to Y)$ where $G$ is a topological group; besides given
the Hilbert inner product on a fiber $\pi^{-1}(p)$ and an invariant measure
class $\mu$ we can introduce 
$$\mathcal{H}=\left\{\psi\mid \psi\; {\textrm\; a\; Borel\; section\; of\; the\; bundle,}
\int\limits_Y<\psi(p),\psi(p)>d\mu(p) < \infty\right\}.$$

A unitary G-action on $\mathcal{H}$ is given by:
$$(g\psi)(p)=\sqrt{\frac{d\mu_g}{d\mu}(p)}g(\psi(g^{-1}\cdot p)),\;\;\;\forall
p\in Y.$$
This action is also a representation of $G$ on $\mathcal{H}$ and does not
 depend on the measure $\mu$. Moreover this construction grants us that
for any G-Hilbert bundle $\zeta_\sigma=(\pi:G\times_K V\to G/K)$ defined by a
locally compact group $G$ and by a K-representation $\sigma$, it is possible
 to derive an {\it induced representation}
$T(\zeta_\sigma)$ which basically tells us that from any representation $\sigma$
of $K$ we can de facto induce a representation $T$ to $G$. 
Consider now a group $G$ which contains an
abelian normal subgroup $N$. If we choose a subgroup $H$ such that the map
$N\times H$ is bijective, we can show that there exists an isomorphism between G
and the semidirect product of $N\ltimes H$.

A character of $N$ is a continuous homomorphism
\begin{equation}
\chi:N\longrightarrow U(1).
\end{equation}
The set of all these maps forms an abelian group called the {\it dual group}:
$$\hat{N}=\left\{\chi\mid (\chi_1\chi_2)(n)=\chi_1(n)\chi_2(n)\right\}$$
Define then a G-action ($G\sim N\ltimes H$) onto the dual space
induced from $G\times N\to N$ letting $(g,n)\to g^{-1}ng$ such that
$$G\times\hat{N}\to\hat{N}$$
gives $g\chi(n)=\chi(g^{-1}ng)$. Thus for any element $\chi\in\hat{N}$, one can
define the {\it orbit} of the character as:
$$G\chi=\left\{g\chi\;\mid\; g\in G\right\},$$
and the {\it isotropy group} of a character under the G-action as:
$$G_\chi=\left\{g\; \mid\; g\in G,\;\;g\chi=\chi\right\}.$$
Clearly the set $G_\chi$ is never empty due to the fact that $N$ acts trivially
onto a character. Introduce now $L_\chi=H\cap G_\chi$, then
$$G_\chi=N\ltimes L_\chi.$$
The group $L_\chi$ is called the {\it little group} of $\chi$ and it is the
isotropy group of the character $\chi$ under the action of the subgroup $H\subset
G$.

Consider now a unitary representation $\sigma$ for the little group $L_\chi$ acting on a
vector space $V$. Then the map
$$\chi\sigma: N\times V\to U(N\times V),$$ 
such that $(n,v)\to \chi(n)\sigma(v)$, is a unitary representation of
$G_\chi$ on the vector space $V$. So one can
introduce an Hilbert G-bundle $\pi:G\times_{G_\chi} V\to G/G_\chi$ with a base
space isomorphic to the space of orbits $G\chi$ defined for every representation
$\chi\sigma$. One finally has \cite{mackey}: 
\begin{theorem}[Mackey]
Let $G=N\ltimes H$ be a semidirect group as above and suppose that $\hat{N}$
contains a Borel subset meeting each orbit in $\hat{N}$ in just one point. Then
\begin{itemize}
\item The representation $T(\zeta)$ induced by the bundle $\pi:G\times_{G_\chi}
V\to G/G_\chi$ is an irrep of $G\sim N\ltimes H$ for any $\chi$ and for any
$\sigma$.
\item each irrep of $G$ is equivalent to a representation $T(\zeta)$ as above
with the orbit $G\chi$ uniquely determined and $\sigma$ determined only up to
equivalence.
\end{itemize}  
\end{theorem}
\subsection{BMS representations in Hilbert topology}

We endow the supertranslation group
with an Hilbert inner product:
\begin{equation}
<\alpha,\beta>=\int\limits_{S^2}\alpha(x)\beta(x)d\Omega;
\end{equation}
where $x\in S^2$, and the supertranslations $\alpha,\beta$ are scalar maps
$S^2\to\Re$.
Therefore $N=L^2(S^2)$ is an abelian
topological group. 

Any element $\alpha$ in the supertranslation group can be decomposed as:
$$\alpha(\theta,\phi)=\sum\limits_{l,m}\alpha_{lm}Y_{lm}(\theta,\phi).$$
This decomposition is topology independent but in the case we are
 considering the complex coefficients $\alpha_{lm}$
have to satisfy
$$\bar{\alpha}_{lm}=(-)^m\alpha_{l,-m}.$$
Notice that supertranslations
admit a natural decomposition into the direct sum of two orthogonal (under the
Hilbert space internal product) subspaces (i.e. subgroups): translations and
proper supertranslations. In particular, for any $\alpha(\theta,\phi)\in
L^2(S^2)$, one can write $\alpha=\alpha_0+\alpha_1$ with:
$$\alpha_0=\sum\limits_{l=0}^1\sum\limits_{m=-l}^l\alpha_{lm}Y_{lm}(\theta,\phi),\;\;\;\alpha_1=\sum\limits_{l>1}\sum\limits_{m=-l}^l\alpha_{lm}Y_{lm}(\theta,\phi).$$
Thus $N$ can be written as:
$$N\sim A\oplus B,$$
where $A$ is the translation group and $B=N-A$. One has however to keep in mind
that this decomposition, as an isomorphism, is not preserved under the action of the
$SL(2,\mathbb{C})$ group. 

Consider the dual of the supertranslation space, the character
space $\hat{N}$ whose elements can be written exploiting the Reisz-Fischer theorem for
Hilbert spaces as: 
$$\chi\in\hat{N}\Longrightarrow \chi(\alpha)=e^{i<\phi,\alpha>},$$
where $\phi\in N$ is uniquely determined. The G-action on $\hat{N}$ is defined as
the map $G\times\hat{N}\to\hat{N}$ sending the pair $(g,\chi)$ to
$g\chi(\alpha)=\chi(g^{-1}(\alpha))$; instead from the point of view of an
element $\phi\in N$,
the action $G\times N\to N$ is:
$$g\phi(z,\bar{z})=K_g^{-3}(z,\bar{z})\phi(gz,g\bar{z}).$$

The above relation tells us that the
dual space $\hat{N}$ is isomorphic to the supertranslations space
 $N$ and there exists a decomposition of $\hat{N}$ as a direct sum of two 
 subgroups-i.e.
$\hat{N}=A_0\oplus B_0$, where $A_0$ is (isomorphic to) the space of linear functionals   
vanishing on $A$ whereas $B_0$ is the space of linear functionals vanishing on
$B_0$.
This means that $A_0$ is composed by those character
mapping all the elements of $A$ into the unit number and the same holds also for $B_0$.
As in the supertranslation case, this decomposition is only true at the level of vector
fields since it is not G-invariant. The only space which is not changing under group
transformations is the subspace $A_0$.

Since one can associate a unique element of $N$, namely $\phi$ to each $\chi(\alpha)$ one 
can
also decompose this field as:
$$\phi(\theta,\phi)=\sum\limits_{l=0}^1\sum\limits_{m=-l}^l
p_{lm}Y_{lm}(\theta,\phi)+\sum\limits_{l>1}\sum\limits_{m=-l}^l
p_{lm}Y_{lm}(\theta,\phi),$$
where the first piece in the sum is in one to one correspondence with the quadruplet
$(p_0, p_1, p_2, p_3)$ which can be thought as the components of a momentum vector
related to the Poincar\'e group. For this reason one can introduce a new space
$A^\prime$ isomorphic both to $A$ and $\hat{A}$ which is given by the set of all
possible functions $\phi$ and which is often referred as the supermomentum space.

One can now find representations of the BMS group in Hilbert topology with the
help of Mackey's theorem applying it to this infinite dimensional (Hilbert)-Lie
group in the spirit of \cite{Mc1}. The first step consists in finding the orbits of $SL(2,\mathbb{C})$ in
$\hat{N}$ which are homogeneous spaces that can be classified as the elements of the set of non
conjugate subgroups of $SL(2,\mathbb{C})$. In order to find a representation for the BMS
group, after classifying the homogeneous spaces $M$, we shall find a character $\chi_0$
fixed under $M$ and then identify each $M$ with its little group associated with the
orbit $G\chi_0\sim G/L$. 

As a starting point we shall consider only connected subgroups of
$SL(2,\mathbb{C})$. The list of these groups is well known but most of
them do not admit a non trivial fixed point in $N$. This request restricts them
to

\begin{center}
\begin{tabular}{|c|c|c|c|}
\hline
Little group & Character & Fixed point & Orbit\\
\hline
$SU(2)$ & $\chi(\alpha)=e^{i<K,\alpha>}$ & $\phi(\theta,\varphi)=K$ &
$PSL(2,\mathbb{C})$\\
\hline
$\Gamma$ & $\chi(\alpha)=e^{i<\zeta(\mid
z\mid),\alpha>}$ & $\zeta(\mid z\mid)$ & $G/\Gamma$\\
\hline
$Z_2$ & $\chi(\alpha)=e^{i<\phi_0(z,\bar{z}),\alpha>}$ & $\phi_0(z,\bar{z})$ &
$G/SU(2)$\\
\hline
\end{tabular}
\end{center}

\noindent where $\Gamma$, which consists of diagonal matrices, is the double cover of
$SO(2)$ and where $Z_2$ is not formally a connected group but nonetheless it is the center of
$SL(2,\mathbb{C})$ and for this reason it acts in a trivial way.

At this point one needs to express explicitly the induced representations;
this operation consists in giving a unitary irrep $U$ of $L_\chi$ on a suitable
Hilbert space $H$ for any little group and a G-invariant measure on the orbit of
each little group. For the connected subgroups, one has \cite{Mc1}:
\begin{itemize}
\item the group $Z_2$ has only two unitary irreps, the identity $D^0$ and a second
faithful representation $D^1$ both acting on the Hilbert space of complex numbers
$\mathbb{C}$ as:
$$D^0(\pm I)=1,\;\;\;D^1(\pm I)=\pm 1.$$
\item the unitary irreps of $\Gamma(\sim \pi\frac{R}{4}Z)$ are instead indexed
by an integer or half integer number $s$ acting on the Hilbert space of complex numbers
$\mathbb{C}$ as:
$$D^s(g)=e^{ist},$$
where $g\in\Gamma$ and
$$g=\left[\begin{array}{cc}
e^{\frac{it}{2}} & 0\\
0 &  e^{\frac{-it}{2}}
\end{array}\right].$$ 
\item the unitary irreps of $SU(2)$ are the usual ones acting on a $2j+1$
dimensional complex Hilbert space with $j\in\frac{Z}{2}$.
\end{itemize}

Consider now the case of non connected little
groups; the hope is that all these groups are compact since this
grants us that their representations can be labelled only by
finite indices. For the BMS group in the Hilbert
topology this is indeed the case since it was shown in
\cite{Mc2} (see theorem 1) that all little groups are compact. 
Besides, since the homogeneous Lorentz
group admits $SO(3)$ as maximal compact subroup, we need to analyse
only subgroups of $SO(3)$. The list of these subgroups has been in \cite{Mc2}:
$$ C_n\;\;\; D_n\;\;\; T\sim A_4\;\;\; O\sim S_4\;\;\; I\sim
A_5\;\;\; \Theta=\Gamma R_2,$$
where $R_2=\left\{I, J=\left[\begin{array}{cc}
0 &1\\
-1 & 0
\end{array}\right]\right\}$. It shows that we are dealing only with groups with finite dimensional
representations which means that there are no continuous indices
labelling states invariant under BMS group with Hilbert
topology.

As in the usual approach, one can then construct on each orbit a
constant function, namely the Casimir, in order to classify them.
Instead of working on $\hat{N}$ it is easier to consider the space
of scalar functions $N^\prime$ isomorphic to $\hat{N}$ and endow
it with a bilinear application assigning to the pair $(\phi_1,\phi_2)$ the number
$B(\phi_1,\phi_2)=\pi(\phi_1)\cdot\pi(\phi_2)$ where $\pi$ is the
projection on the momentum components (i.e. $\pi:N^\prime\to
A^\prime$) and the dot denotes the usual Lorentz inner product. It
is also straightforward to see that the bilinear application is
G-invariant.

This last property implies that on each orbit in $\hat{N}$, the
function $B$ is constant and its value can be calculated since
$\pi(\phi)=(p_0,p_1,p_2,p_3)$ so that:
$$B(\phi,\phi)=\pi(\phi)\cdot \pi(\phi)=m^2. $$

One can thus label each orbit in the
character space with an invariant, the squared mass, together with
the sign of the ``temporal'' component i.e. $sgn(p_0)$. These invariants
grant only a partial classification since, for example, in the case
of unfaithful representations, $\pi(\phi)=0$, which implies
that the above invariants are trivial. For faithful
representations too, we cannot conclude that the classification is
complete since different orbits can correspond to the same value
for the mass. One can also find a constant
number to label the orbits corresponding to unfaithful
representations-i.e. an bilinear invariant application mapping at
least $A^0$ to real numbers. This has been done in
\cite{Mc2}:
\begin{equation}
Q^2=\pi^2\int\int\frac{\mid z_1-z_2\mid^2\ln\mid z_1-z_2\mid}{(1+\mid
z_1\mid^2)(1+\mid
z_2\mid^2)}\phi(z_1,\bar{z}_1)\phi(z_2,\bar{z}_2)d\mu(z_1,\bar{z}_1)d\mu(z_2,\bar{z}_2),
\end{equation}
where $\phi$ is a function of class $C^\infty(S^2)$. Thus $Q^2$ is
defined only for a subset dense in the Hilbert space $L^2(S^2)$
which from the physical 
point of view makes no difference.
\subsection{BMS representations in nuclear topology}
The study of BMS group to label elementary particles started with
 the hope to remove the difficulty  with 
 the Poincar\'e group concerning the continuous representations
associated to the non compact $E(2)$ subgroup. Unfortunately, continuous
representations appear if one chooses
for the supertranslations a different topology (for istance $C^k(S^2))$. This was for
the first time pointed out in \cite{Girardello} where
it was shown that also non compact little
groups appear (for instance $E(2)$).  

Nonetheless it is worth studying representations for the BMS group with $N$
endowed with the topology $C^\infty(S^2)$. In this case the action of $SL(2,\mathbb{C})$ on the space of
supertranslations is given by a representation $T$ equivalent to the irrep of $SL(2,\mathbb{C})$
on the space $D_{(2,2)}$ introduced by Gel'fand. This implies that we have to
use techniques proper of {\it rigged} Hilbert spaces. 

The main object we shall deal
with is $D_{(n,n)}$ which is the space of functions $f(z,\omega)$ of
class $C^\infty$ except at most in the origin. These functions also satisfy the relation $f(\sigma
z,\sigma w)=\mid\sigma\mid^{(2n-2)}f(z,w)$ for any $\sigma\in\mathbb{C}$. At the
end of the day, one has the following chain of isomorphisms
$$BMS=N\ltimes G\longleftrightarrow D_{(2,2)}\ltimes G\longleftrightarrow D_{2}\ltimes
G,$$
where $D_2$ is the space of $C^\infty$ functions $\zeta(z)$ depending on a
single complex variable such that any element $g(z,w)\in D_{(2,2)}$ can be
written as $g(z,w)=\;\;\mid z\mid^{2}\zeta(z_1)=\mid w\mid^2\hat{\zeta}(z_1)$
with $z_1=\frac{w}{z}$ and $\hat{\zeta}(z_1)=\mid z_1\mid^2\zeta(z_1^{-1})$.

Irreducible representation can arise (see theorem 2 in \cite{Mc4}) can arise either from a transitive G action in the
supermomentum space or from a cylinder measure $\mu$ with respect to the G
action is strictly ergodic i.e. for every measurable set $X\subset N^\prime$ $\mu(X)=0$ or
$\mu(N^\prime-X)=0$ and $\mu$ is not concentrated on a single G-orbit in
$N^\prime$. 

The first step is to classify all little groups; they can either
be discrete subgroups, non-connected non discrete Lie subgroups and
connected Lie subgroups. 

Discrete subgroups can be derived exactly as in the Hilbert
case and so the only connected little groups for the BMS group are:
$$SU(2),\;\;\Gamma,\;\;\Delta,\;\;SL(2,\mathbb{R}).$$

Here one can point out the first difference between the Hilbert and the nuclear
topology which consists in the appearance of the $SL(2,\mathbb{R})$ little group
which will contribute only to unfaithful representations.

Non connected non discrete subgroups $S$ can be derived using still theorem
5 in \cite{Mc4} since each $S$ is a subgroup of the normalizer $N(S_0)$ where
$S_0$ is the identity component of $S$. Here is the list:
\begin{itemize} 
\item $S_1$ which is the set of matrices $$\left\{\left[\begin{array}{cc}
\sigma^r & 0\\
0 & \sigma^r\end{array}\right] {\rm with}\;\; \sigma=e^{\frac{2\pi i}{n}},\;\;0\leq
r\leq(n-1)\right\},$$ 
\item $S_2$ which is the set of matrices $$\left\{\left[\begin{array}{cc}
e^{qr} & 0\\
0 & e^{-qr}\end{array}\right] \textrm{where q is a fixed non negative number and r
is an integer}\right\},$$
\item $S_3$ which is the set of matrices $$\left\{\left[\begin{array}{cc}
z_1^{r}z_2^s & 0\\
0 & z_1^{-r}z_2^{-s}\end{array}\right] {\rm where}\; z_1, z_2\in\mathbb{C}\;
\textrm{and r, s are integers},\right\},$$
\item $S_4$ which is the set of matrices $$\left[\begin{array}{cc}
1 & r\\
0 & 1\end{array}\right].$$
\end{itemize}

To establish the faithfulness of the irreps, one needs to calculate the
projection on the supermomentum space of supertranslation:
$$\pi(\phi)(z^\prime)=\frac{i}{2}\int
dzd\bar{z}(z-z^\prime)(\bar{z}-\bar{z^\prime})\phi(z)\neq 0.$$

The only connected groups occuring as little
group for faithful representations are $\Gamma,\Delta,SU(2)$ with vanishing
square mass $0$. It is also interesting to
notice that the orbit invariant $Q^2$ defined for the Hilbert topology for
unfaithful representations is not available in the nuclear topology since it is
not defined for distributional supermomenta. Finally no information is available about discrete subgroups
since it very difficult to classify them above all for infinite discrete
subgroups. This means that the study of BMS representations in the nuclear
topology has to be completed yet.
\section{Wave equations in fiber bundle approach}
We are going to briefly review in the following definitions and notations 
used in the derivations of the BMS wave equations following \cite{Asorey}. 
As said we use sort of diagrams to facilitate the reader even if they are
not rigorous mathematically speaking.

Consider then
$$\xymatrix{P(H,M)\ar@{-->}[r]^{\sigma}\ar[d]^{\pi} & GL(V)\ar@{-->}[d]^{*} \\
M & E(M,V)\ar[l]^{\pi_E}}$$
In our  case $P(H,M)$ is a group $G$ and a principal bundle whose
fiber $H$ is a closed subgroup of $G$ and whose base space is the homogeneous space
given by the coset $G/H$;
each linear representation $\sigma:H\to GL(V)$ automatically defines the vector bundle
(that's the reason why we used the dotted lines) $E(M,V)=P\times_H V$ whose generic
element is the equivalence class $[u,a]$ with $u\in P$ and $a\in V$. The equivalence
relation defining this class is given by
$(u,a)\sim(uu^\prime,a)=(u,\sigma(u^\prime)a)$. 

One defines than a G-action; in particular on
$P=G$ it is the obvious one i.e.
$g(uh)=(gu)h$ whereas on $M$ the action is induced through the projection $\pi$ as
$g\pi(u)=\pi(gu)$. Finally on the E bundle the G-action is $g[u,a]=[gu,a]$. i
if we consider a generic section for the E bundle i.e. $\psi:M\to E(M,V)$, we can act
on it through a linear representation of G as:
\begin{equation}\label{psi1}
\left(U(g)\psi\right)(gm)=g\psi(m).
\end{equation}
This representation is exactly the ``induced'' representation of $G$ constructed from
the given one $\sigma$ of the subgroup $H$.
Moreover if $\sigma$ is pseudo-unitary and it exists an invariant G-measure on M we can
define an internal product $(,):\Gamma(E)\times\Gamma(E)\to\mathbb{C}$ as:
$$<\psi,\phi>=\int\limits_M d\mu h_m(\psi,\phi),$$
where $h_m$ is the induced internal product on the fibre $\pi^{-1}_E(x)$.

In the specific case of semidirect product of groups i.e. $G=N\ltimes K$ (N
abelian), one can define a G-action on the character space $\hat{N}$:
$$g\chi(gn)=\chi(n).$$

For any element $\chi_0$ one can construct its stability (little) group
$G_{\chi_0}:N\ltimes K_{\chi_0}$ and assign a representation $\sigma:K_{\chi_0}\to
GL(V)$. This induces a representation $\chi_0\sigma:G_{\chi_0}\to V$ which 
associates with the
couple (n,g) the element $\chi_0(n)\sigma(g)$. Thus in our diagram the group
$G_{\chi_0}$ is playing the role of $H$ and $M$ becomes the coset space
$G/G_{\chi_0}$. The diagram is then: 
$$\xymatrix{G(N\ltimes K_{\chi_0}, G/G_{\chi_0})\ar@{-->}[r]^{\:\:\;\;\;\;\;\;\;\;\chi_0\sigma}\ar[d]^{\pi} & GL(V)\ar@{-->}[d]^{*} \\
G/G_{\chi_0} & E(G/G_{\chi_0},V)\ar[l]^{\pi_E}}$$
Since we want to describe vector valued functions $f:G/G_{\chi_0}\to V$, a representation of  
$U$ can be made fixing a section $s:G/G_{\chi_0}\to G$ and remembering that for any
element $\psi\in\Gamma(E)$ there exists a function $\tilde{f}_{\psi}:P\to V$:
$$\psi(\pi(u))=[u,\tilde{f}_\psi(u)].$$
A vector valued function is:
$$ f_\psi=\tilde{f}_\psi\circ s.$$

Let us notice that this construction makes sense only if the section $s$ is global
otherwise $f$ is not defined everywhere; this happens only if the bundle $G$ is
trivial i.e. $G=M\times H$ which is always the case in the situation we are interested
in.

Moreover equation (\ref{psi1}) translates in:
\begin{equation}\label{psi2}
\left[U(g)f_\psi\right](gx)=\sigma(\gamma(g,x))f_\psi(x),
\end{equation}
where in our case $\gamma:(N\ltimes K)\times G/G_{\chi_0}\to G_{\chi_0}$ is defined
as:
\begin{equation}\label{gamma}
s(gx)\gamma(g,x)=gs(x)
\end{equation}
From now on we shall call (\ref{psi2}) the \emph{canonical (or induced) wave
equation}.

In physical relevant situations covariant representations are used
instead of induced ones. In this case we deal with a principal bundle
$G(X,G_{x_0})$ where $x_0\in X$ and the interesting representations are those
preserving locality on the physical relevant space $X$ and acting on a vector
valued function $f:X\to V$ as:
\begin{equation}\label{psi3}
[T(g)f](gx)=A(g,x)f(x),
\end{equation}
where $A$ is a map from $G\times X$ to $GL(V)$ satisfying the property:
$$A(g_1g_2,x)=A(g_1,g_2x)A(g_2,x).$$
Examples of these representations are those induced from the isotropy group
$G_{x_0}$ when expressed in term of sections $s:X\to G$. Let us also notice that
there exists a map $\Sigma:G_{x_0}\to GL(V)$ assigning to an element $\gamma$
the matrix $\Sigma(\gamma)=A(\gamma,x_0)$ and the induction of such a
representation from the isotropy group to the entire $G$ generates the
representation
$$A^\prime(g,x)=\Sigma(\gamma(g,x)),$$
where $\gamma$ is defined as in (\ref{gamma}) and $g=(\phi,\Lambda)$.


\end{document}